%% file: arxiv_main.tex
\documentclass[aps,prl,reprint,superscriptaddress,amsmath,amssymb,
               floatfix,nofootinbib,longbibliography]{revtex4-2}

\pdftrailerid{}

\usepackage{graphicx}
\usepackage{bm}
\usepackage{mathtools}
\usepackage{amsthm}
\usepackage{xcolor}
\usepackage{enumitem}
\usepackage{hyperref}
\hypersetup{
  colorlinks=true,
  linkcolor=blue!55!black,
  citecolor=blue!55!black,
  urlcolor=blue!55!black,
}

\newcommand{\TB}{\mathrm{TB}}
\newcommand{\BP}{\mathrm{BP}}

\newcommand{\non}{\mathrm{on}}
\newcommand{\noff}{\mathrm{off}}
\newcommand{\chiralcur}{\mathrm{chiral}}
\newcommand{\as}{\mathrm{as}}

\newcommand{\vj}{\mathbf{j}}
\newcommand{\vl}{\mathbf{l}}
\newcommand{\vk}{\mathbf{k}}

\newcommand{\parahead}[1]{\textit{#1.}---}

\begin{document}

\title{Anomalous Pairing Currents and a Second Topological Edge Channel \\ in Bosonic Lattices}

\author{Chitrak Bhadra}
\affiliation{Departamento de F\'{i}sica Te\'orica, Universidad
Complutense de Madrid, 28040 Madrid, Spain}

\author{\'Angel Rivas}
\affiliation{Departamento de F\'{i}sica Te\'orica, Universidad
Complutense de Madrid, 28040 Madrid, Spain}
\affiliation{CCS-Center for Computational Simulation, Campus de
Montegancedo UPM, 28660 Boadilla del Monte, Madrid, Spain}

\author{Miguel A.\ Mart\'{i}n-Delgado}
\affiliation{Departamento de F\'{i}sica Te\'orica, Universidad
Complutense de Madrid, 28040 Madrid, Spain}
\affiliation{CCS-Center for Computational Simulation, Campus de
Montegancedo UPM, 28660 Boadilla del Monte, Madrid, Spain}

\date{\today}

\begin{abstract}
We show that bosonic pairing opens a second chiral current channel on a 2D kagome lattice, absent from any particle-conserving model. From the continuity equation, we derive both a hopping current and an anomalous pairing current on the lattice, and predict chiral circulation in bulk-gapped phases with integer para-unitary Chern numbers, as well as a phase-sensitive leakage ratio, $\Lambda_{\cal I}$, for the pairing current around a defect. This ratio can be tuned from confined to strongly anomalous regimes at fixed topology. The two channels differ microscopically: the hopping current is sourced by the on-bond single-particle coherence, whereas the pairing current is sourced by the off-site anomalous coherence, whose spatial range is governed by the BdG pairing gap rather than by the single-particle gap. This separation produces distinct defect-induced signatures in real space, identifies a regime in which bulk topology and anomalous edge response coexist with no analogue in particle-conserving matter, and is directly testable in driven photonic lattices and superconducting-circuit arrays.
\end{abstract}

\maketitle

\makeatletter
\let\orig@addcontentsline\addcontentsline
\renewcommand{\addcontentsline}[3]{}
\makeatother

\input{main_body.tex}

\input{arxiv_main_refs.bbl}
\clearpage
\onecolumngrid

\makeatletter
\let\addcontentsline\orig@addcontentsline
\makeatother

\setcounter{section}{0}
\setcounter{subsection}{0}
\setcounter{equation}{0}
\setcounter{figure}{0}
\setcounter{table}{0}
\setcounter{page}{1}
\renewcommand{\thesection}{S\Roman{section}}
\renewcommand{\thesubsection}{\thesection.\arabic{subsection}}
\renewcommand{\thefigure}{S\arabic{figure}}
\renewcommand{\theequation}{S\arabic{equation}}
\renewcommand{\thetable}{S\arabic{table}}
\setcounter{secnumdepth}{3}
\makeatletter
\renewcommand\p@subsection{}
\makeatother

\begin{center}
{\large\bfseries Supplemental Material:\\[2pt]
Anomalous Pairing Currents and a Second Topological Edge Channel \\
in Bosonic Lattices}\\[12pt]
{Chitrak Bhadra,$^{1}$ \'Angel Rivas,$^{1,2}$ and Miguel A.\ Mart\'{i}n-Delgado$^{1,2}$}\\[6pt]
{\small\itshape $^{1}$Departamento de F\'{i}sica Te\'orica, Universidad Complutense de Madrid, 28040 Madrid, Spain}\\
{\small\itshape $^{2}$CCS-Center for Computational Simulation, Campus de Montegancedo UPM, 28660 Boadilla del Monte, Madrid, Spain}\\[6pt]
{\small(Dated: \today)}
\end{center}

\vspace{1em}

\makeatletter
\renewcommand{\l@section}[2]{%
  \ifnum \c@tocdepth >\z@
    \addpenalty\@secpenalty
    \addvspace{1.0em \@plus\p@}%
    \setlength\@tempdima{3.0em}%
    \begingroup
      \parindent \z@ \rightskip \@pnumwidth
      \parfillskip -\@pnumwidth
      \leavevmode \bfseries
      \advance\leftskip\@tempdima
      \hskip -\leftskip
      #1\nobreak\hfil \nobreak\hb@xt@\@pnumwidth{\hss #2}\par
    \endgroup
  \fi}
\renewcommand{\l@subsection}{\@dottedtocline{2}{3.0em}{3.2em}}
\makeatother

\addtocontents{toc}{\protect\setcounter{tocdepth}{2}}
\providecommand{\contentsname}{Contents}
\tableofcontents

\vspace{1em}

\makeatletter
\renewcommand\@biblabel[1]{[S#1]}
\renewcommand{\citenumfont}[1]{S#1}
\makeatother

\input{sm_body.tex}

\input{arxiv_sm_refs.bbl}
\end{document}

%% file: main_body.tex
\parahead{Introduction}%
Chiral edge transport is the hallmark of two-dimensional topological
phases \cite{HasanKane2010,QiZhang2011,Chiu2016}, and its microscopic origin
in particle-conserving lattices is by now well understood through
Chern-number arguments and the bulk-boundary correspondence. Bosonic systems with anomalous pairing
terms, however, require a separate treatment: the Bogoliubov--de
Gennes (BdG) Hamiltonian remains Hermitian, but its diagonalization
must preserve the bosonic commutation algebra and therefore proceeds
by a para-unitary transformation with respect to the paravector
metric $\Sigma_{z}=\operatorname{diag}(\mathbf{1},-\mathbf{1})$,
rather than by an ordinary unitary rotation
\cite{Colpa1978,Shindou2013,SM}. As a direct consequence, the lattice continuity equation
splits the particle current into two physically distinct
contributions---a conventional hopping single-particle current (SPC) and an anomalous pairing current (PC).  Despite growing interest in
pairing-induced topology in photonic and circuit-QED
platforms \cite{Peano2016,Chaudhary2021}, a sharp statement of what
pairing actually adds to the edge sector, at the level of the
microscopic currents themselves, has been missing.

\begin{figure}[t]
  \centering
  \includegraphics[width=\columnwidth]{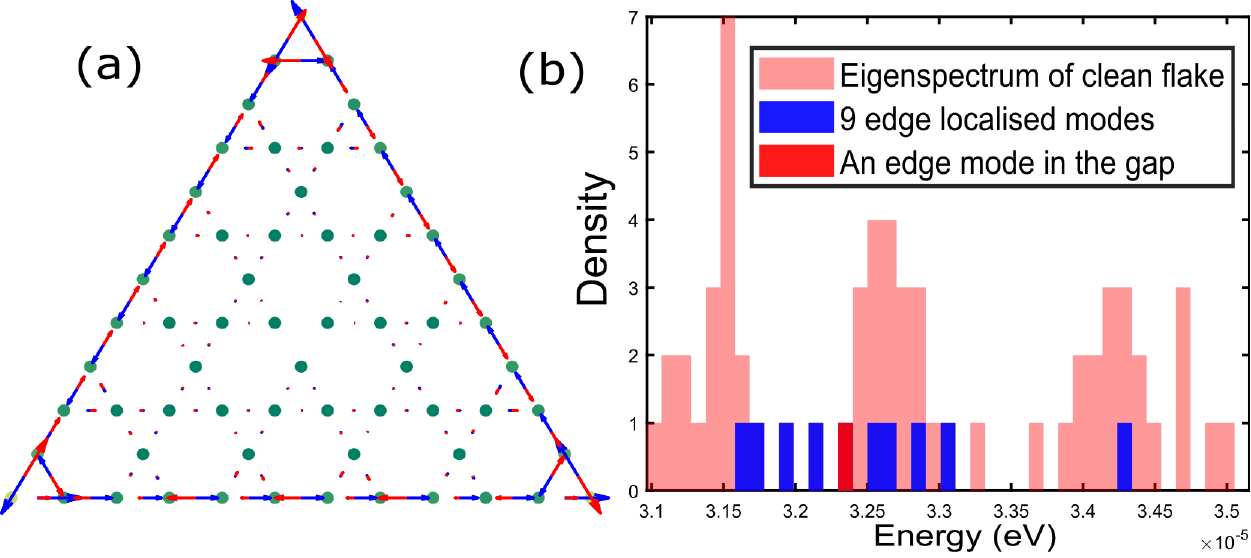}
  \caption{Microscopic decomposition of the chiral edge transport in a 2D bosonic kagome lattice on a finite zigzag-terminated flake.  In (a),  blue/red arrows indicate Single Particle Current/Pairing Current for non-zero particle-conserving as well as bosonic pairing interactions. In (b) the eigenspectrum is shown with red indicating an edge localized topologically protected mode. Blue indicates nine other edge localized modes of varying gap/bulk features. Parameters: $\omega_{0}=50$, $J=0.02\,\omega_{0}$,   $\nu_{\non}=3J$, $\nu_{\noff}=0.1J$ (Lattice size $N=63$); these values are chosen so as to produce opposing chirality of SPC and PC on the same edge.}
  \label{fig:lattice-edge}
\end{figure}

In this Letter we establish such a statement.  We show that the new PC enabled by bosonic pairing represents a different topologically robust and chiral transport channel,  with no analogue in particle-conserving models. In addition, we show the presence of anomalous defect-scattering signature that can be tuned via the phase structure of the Hamiltonian.   We realize this mechanism in a 2D kagome lattice with coexisting hopping and pairing terms. This provides a minimal setting in which both sectors can be phase-tuned independently on the same geometry.

On this platform, we derive the two lattice current operators, compute them in real space around localized edge and bulk defects,  and connect the chirality of the observed current patterns to integer para-unitary Chern numbers of the bulk bands. The fermionic bulk--boundary correspondence does not transfer literally to this bosonic setting. The mechanism is structural: with $m$ the number of physical bands per unit cell of the lattice (so that the BdG matrix is of size $2m\times 2m$), the para-unitary diagonalization of the bosonic Bogoliubov Hamiltonian has structure group $U(m,m)$ rather than the compact $U(2m)$ of the fermionic BdG case; the index theorem that ties the bulk Chern integer to a count of protected chiral edge channels relies on the compactness of $U(2m)$ and on the positive-definiteness of the bundle metric, neither of which holds for $U(m,m)$. Finite-size scaling on ribbons up to $L=64$ (see~\cite{SM}, Sec.~V.1) confirms that this is not a finite-resolution artefact: the bulk Chern numbers are accordingly to be read as phase-sensitive observables of the Bogoliubov bundle rather than as strict counters of edge modes (see~\cite{SM}, Sec.~V).
Our primary evidence for non-trivial chiral transport is therefore furnished by fully non-perturbative simulations on finite zigzag flakes, whose robustness is reinforced by Chern stability under parameter variations and by quantifying  phase-sensitive leakage of currents away from the immediate neighborhood of the defect sites. An example of an edge mode on a clean defect-less flake and its position in the gap of the corresponding particle eigenspectrum is presented in Fig.~\ref{fig:lattice-edge}.

Thus, the present work extends two complementary approaches to bosonic pairing topology. While momentum-space band analysis \cite{Peano2016,Chaudhary2021} establishes the existence of topological phases in pairing Hamiltonians, our real-space approach provides a microscopic characterization of the transport phenomena arising from the interplay between tight-binding and pairing sectors.  In particular, it reveals the emergence of distinct anomalous and confined transport regimes, as well as bulk-mediated current-carrying modes.

Complementary real-space approaches to topological transport in bosonic systems have been developed in dissipative and nonequilibrium settings~\cite{Rivas2017,Mitchison2022}; by contrast, the present Letter focuses on the zero-temperature coherent regime, where bosonic pairing alone—without reservoirs or external driving—opens the second topological channel.

\begin{figure*}[!t]
  \centering
  \includegraphics[width=0.95\textwidth]{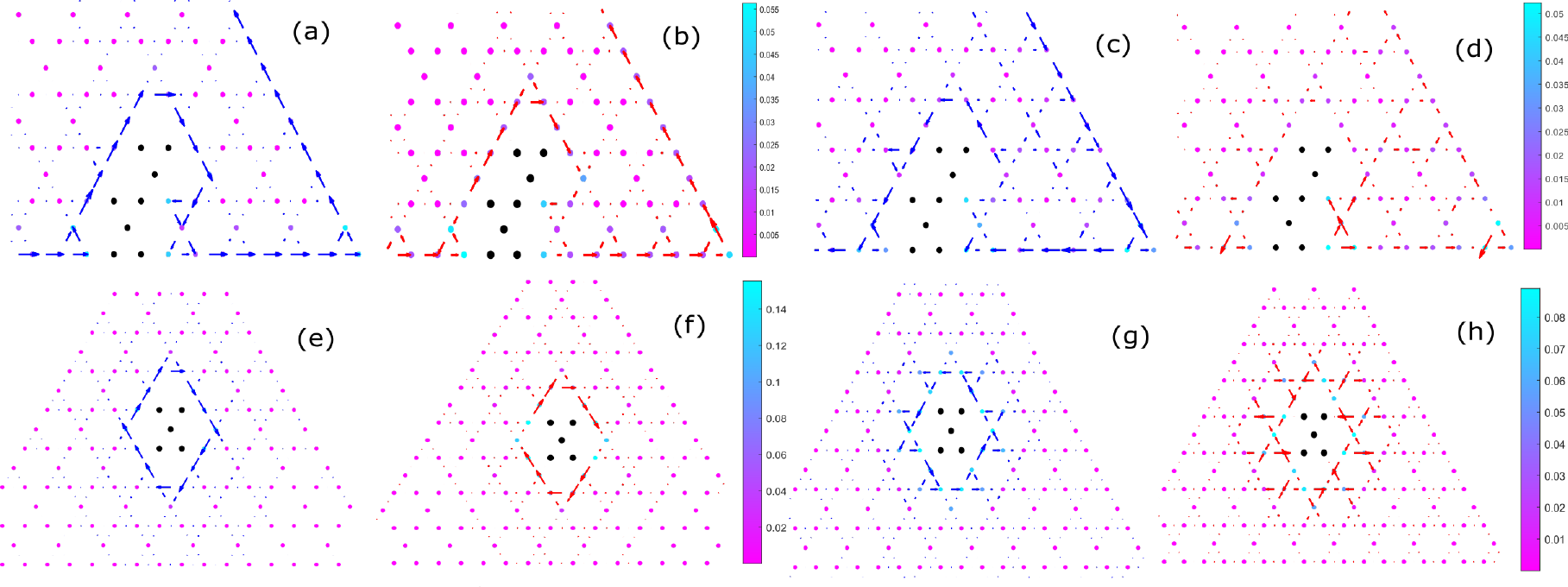}
  \caption{Zoomed view of chiral current decomposition around localized defects in
  the kagome lattice (Lattice size $N=234$).  Columns compare Case~1 with Case~2.  Rows compare a defect at the zigzag edge (top, panels a--d) with a defect deep in the bulk
  (bottom, e--h). Within each pair of columns the left panel shows
  the single-particle-conserving current $\langle \mathcal{J}\rangle$ (blue)
  and the right panel the pairing current
  $\langle \mathcal{I}\rangle$ (red).  Arrows indicate current direction;
  colormap shows the absolute square of the wavefunction at each point.
  The defect sites are marked in black. Parameters: $\omega_{0}=50$,
  $J=0.02\,\omega_{0}$, defect detuning
  $\omega_{\mathcal{S}}=10^{4}\omega_{0}$. Case~1 (a,b,e,f):
  $\nu_{\non}=0.1J$, $\nu_{\noff}=3J$,
  $\psi_{s,s'}=2\pi/3$, $\varphi_{s,s'}=0$. Case~2 (c,d, g,h):
  $\nu_{\non}=0.1J$, $\nu_{\noff}=3J,  \varphi_{AB}=\varphi_{A}=0,\varphi_{BC}=\varphi_{B}=2\pi/3, \varphi_{CA}=\varphi_{C}=4\pi/3$, $\psi=0$.}
  \label{fig:consolidated-defects}
\end{figure*}

\parahead{Kagome lattice model}%
We investigate a bosonic Hamiltonian on a 2D kagome lattice,
\begin{align}
H_{S}
&=\sum_{\vj}\!\left[\omega_{0} a_{\vj}^{\dagger}a_{\vj}-\tfrac{\nu_{\rm on}}{2}\bigl(e^{i\varphi_{s}}a_{\vj}^{\dagger}a_{\vj}^{\dagger}+\mathrm{h.c.}\bigr)\right]\notag\\
&\ -\!\!\sum_{\langle \vj,\vj'\rangle}\!\!\left[J\bigl(e^{i\psi_{s,s'}}a_{\vj}^{\dagger}a_{\vj'}\!+\mathrm{h.c.}\bigr)+\tfrac{\nu_{\noff}}{2}\bigl(e^{i\varphi_{s,s'}}a_{\vj}^{\dagger}a_{\vj'}^{\dagger}\!+\mathrm{h.c.}\bigr)\right]
\label{eq:H}
\end{align}
where $a_{\vj}$ is the bosonic annihilation operator at lattice site
$\vj=(j_{1},j_{2},s)$ with $s=A,B,C$ the sublattice index. The
Hamiltonian splits into a tight-binding (TB) sector with onsite
frequency $\omega_{0}$ and nearest-neighbor hopping $J$ endowed with
a sublattice phase $\psi$, and a Bogoliubov-pairing (BP) sector with
onsite and nearest-neighbor amplitudes $\nu_{\non}$, $\nu_{\noff}$
and phases $\varphi$. Throughout we restrict $J$, $\nu_{\non}$,
$\nu_{\noff}$ to a small fraction of $\omega_{0}$ to keep the system
dynamically stable~\cite{Peano2016}. Para-unitary diagonalization
is performed (details are given
in~\cite{ SM}) and we focus on three flux configurations:\par
\indent - Case~1 ($\psi\neq 0$, $\varphi=0$) threads phases only in the TB sector (Fig.~\ref{fig:lattice-edge} pertains to this case).\par
\indent - Case~2 ($\psi=0$, $\varphi\neq 0$) threads phases only in the BP sector.\par
\indent - Case~3 ($\psi,\varphi\neq 0$) generalizes both and enables flux engineering of the edge transport.

\parahead{Microscopic description of the edge currents}%
The rate of change of the on-site population
$\frac{d}{dt}(a_{\vj}^{\dagger}a_{\vj})$, evaluated via the
Heisenberg equations of motion for the Hamiltonian
\eqref{eq:H}, reveals the simultaneous existence SPC and PC~\cite{SM}:
\begin{align}
 \mathcal{J}_{\vj\vj'}
&=iJ\!\left[a_{\vj}^{\dagger}a_{\vj'}e^{i\psi_{s,s'}}
     -a_{\vj'}^{\dagger}a_{\vj}e^{-i\psi_{s,s'}}\right]
     (1-\delta_{\vj\vj'}),
\label{eq:JSPC}\\
 \mathcal{I}_{\vj\vj'}
&=\frac{i\nu_{\noff}}{2}\!\left[a_{\vj}^{\dagger}a_{\vj'}^{\dagger}
      e^{i\varphi_{s,s'}}-a_{\vj}a_{\vj'}e^{-i\varphi_{s,s'}}\right]
      (1-\delta_{\vj\vj'}).
\label{eq:JPC}
\end{align}
Equations~\eqref{eq:JSPC}--\eqref{eq:JPC} describe the flow of single
and pairs of bosons between neighboring sites, respectively; the on-site
pairing amplitude $\nu_{\non}$ acts as a local sink/source of paired
bosons and modulates the spatial profile of the PC without entering
the current operator directly.

We calculate the expectation values $\langle  \mathcal{J}\rangle\equiv \langle \mathcal{J}_{\vj\vj'}\rangle_E$ and $ \langle\mathcal{I}\rangle\equiv\langle  \mathcal{I}_{\vj\vj'}\rangle_{E}$ for a given eigenmode  of $H_S$ with single particle energy $E$ and wavefunction $\Psi_{\vj,E}$.  Focusing on a localized edge mode (LEM) $\Psi_{\vj,E}^{\rm LEM}$ yields the two currents circulating at the boundary of the lattice (eg. Fig. \ref{fig:lattice-edge}).

\parahead{Edge defects: Case~1 vs.\ Case~2}%
We introduce a connected set of defect sites $\{\mathcal{S}\}$ on one edge of the flake by detuning the on-site energies to $\omega_{\mathcal{S}}=10^{4}\omega_{0}$. They modify the geometric path of a maximally LEM of the corresponding clean flake (see \cite{SM}, Sec. IV.1); consistent with the non-trivial bulk Chern indicators, both currents retain their chirality, but the morphology of the scattering differs markedly between the two cases.

In Case~1 [Fig.~\ref{fig:consolidated-defects}(a)--(b)],  the two currents scatter cleanly around the defect with negligible leakage into neighboring sites. In Case~2 [Fig.~\ref{fig:consolidated-defects}(c)--(d)],  $\langle  \mathcal{J}\rangle$ retains the minimally scattered chiral pattern (the TB sector has no phase), but $\langle  \mathcal{I}\rangle$ still circulates chirally and, crucially,  develops enhanced inhomogeneity around the defect and there is significant leakage into nearest neighbor sites outside the connected loop. This
anomalous scattering pattern is the microscopic fingerprint of the second channel identified in the Introduction.

Three observations underpin the main results of defect-scattering. First, in a purely anomalous configuration with vanishing hopping phase (Case 2), the hopping current $\langle \mathcal{J} \rangle$ circulates chirally along the zigzag edge, while the pairing current $\langle \mathcal{I} \rangle$ is considerably smaller in magnitude. Second, the two currents respond qualitatively differently to localized defects: at an edge defect, $\langle \mathcal{I} \rangle$ develops an anomalous circulation pattern that is negligible in $\langle \mathcal{J} \rangle$; the next subsection reveals that with a bulk defect, \textit{the same asymmetry reappears},  identifying $\langle \mathcal{I} \rangle$ as a genuinely independent transport channel and the microscopic signature of the anomalous-scattering regime introduced below (Figs.~\ref{fig:consolidated-defects} (e)--(h)). Third, tuning the ratio of on-site to off-diagonal pairing and the phases associated with them, drives a crossover between an \textit{anomalous regime}, where the two currents coexist and scatter asymmetrically, and a \textit{confined regime}, where the edge channel is more strongly protected. Together, these microscopic observations identify bosonic pairing as a second, independently tunable topological knob, and frame the defect-scattering signatures discussed below as its experimental fingerprint.

\parahead{Defects in the bulk}%
Topological protection persists when the defect is located in the
bulk.
Eigenmodes exist in both Cases~1 and~2 localized around the bulk defect loop (with different energy than at the flake edge).  The comparison between the two scenarios confirms the diagnostic of the edge-defect analysis: the constant TB phase of Case~1 suppresses anomalous scattering [Fig.~\ref{fig:consolidated-defects}(e)--(f)], while the spatially varying BP phase of Case~2 enhances it [Fig.~\ref{fig:consolidated-defects}(g)--(h)].  Increasingly delocalized $\langle  \mathcal{J}\rangle$ and non-zero $\langle  \mathcal{I}\rangle$ leaking into sites outside the bulk-defect loop markedly differs from a particle-conserving model. Incidentally, for a narrow range in the BP sector of Case~1, there is an eigenmode that carries chiral currents simultaneously along the edge of lattice and bulk defects~\cite{SM}.

\parahead{Clean chiral currents versus anomalous scattering: Case~3}%
We now generalize to Case~3 by allowing complex phases in both
sectors simultaneously, $\psi,\varphi\neq 0$. In the TB sector the
constant phases generate an effective synthetic magnetic flux
through the plaquettes; in the BP sector the spatially varying
phases control the coherence of pair creation and annihilation.
Both phase structures break time-reversal symmetry and together
provide a tuning knob for the morphology of the chiral transport.

\begin{figure}[h]
  \centering
  \includegraphics[width=0.98\columnwidth]{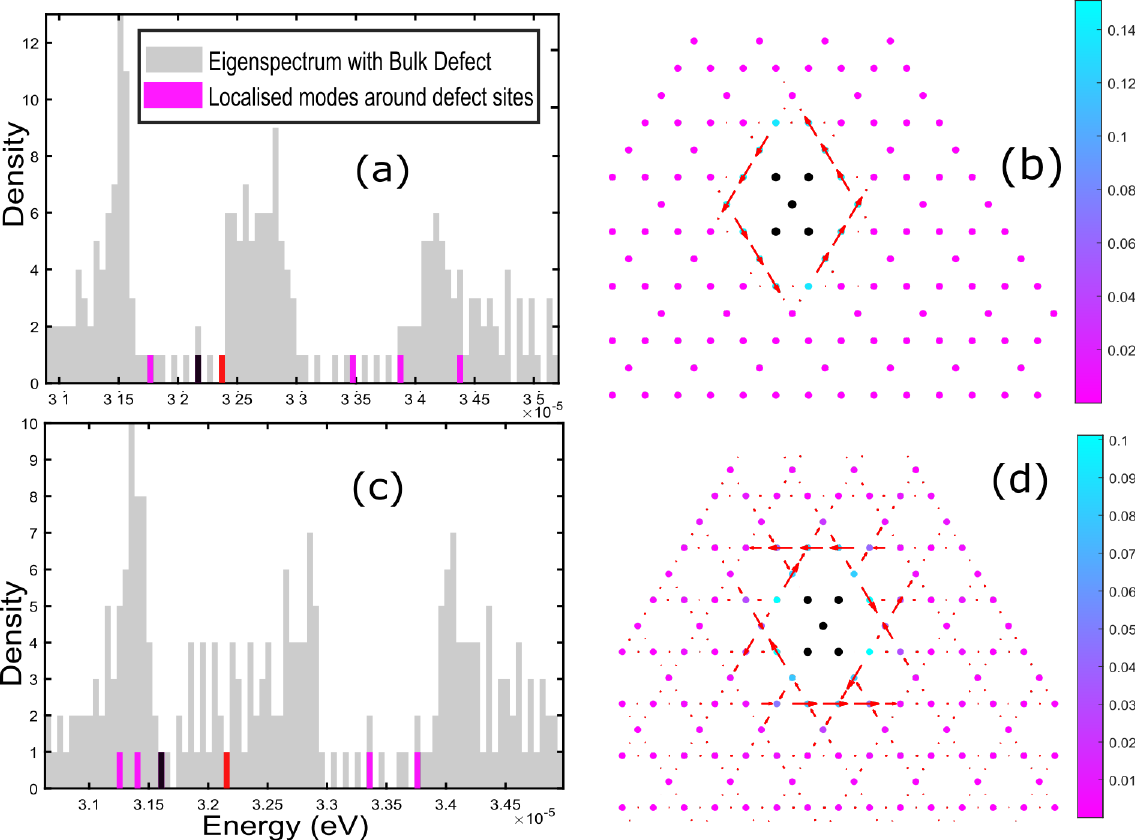}
  \caption{Tuning clean chiral currents against anomalous scattering
  in a flux-engineered Hamiltonian (Case~3, $\psi,\varphi\neq 0$).
  Panels (a-b): $\nu_{\noff}\ll\nu_{\non}$ . Eigenspectrum of the defect-induced localized edge modes shown in (a) with \textbf{black bar} showing the Maximally LEM (MLEM) (also in (c)). Strong localization of currents when defects are introduced in the bulk seen in the example of  $\langle \mathcal{I}\rangle$ shown by red arrows (b).
  Panels (c-d): $\nu_{\noff}\gg\nu_{\non}$. Pairing current $\langle \mathcal{I}\rangle$ around the defect
  loop shows anomalous scattering and
  leakage into neighboring sites shown in (d). In (a) and (c), only the \textbf{red bar} is representative of a localized mode existing on the outer flake boundary, and not on the artificial edge created by the defects.  See a 3D visualization of the corresponding mode densities in section IV.5 of ~\cite{SM}.}
  \label{fig:case3-triple}
\end{figure}

Figure~\ref{fig:case3-triple} illustrates the resulting phenomenology. The eigenspectrum shows that the maximally LEM around bulk defects remains sharply isolated within the bulk gap; meaning, the artificial edge, created in the kagome flake by the defects, invokes topological protection signalling the phenomenon as a true edge effect. Simultaneously, chiral edge modes exist in a different sector of the eigenspectrum that can circulate at the flake boundary with its own distinct edge topology properties.

Remarkably, for a dominant $\nu_{\noff}$ (anomalous regime), the defect mode induced pairing current leaks into neighboring sites,
corroborating the anomalous-scattering fingerprint of
Fig.~\ref{fig:consolidated-defects}(h). For a dominant $\nu_{\non}$
(confined regime), the mode around bulk defects remains sharply localized, the currents wrap cleanly around the loop, and anomalous scattering is
suppressed. In this confined limit the system resembles the particle-conserving hopping model and the anomalous processes are strongly suppressed. We emphasize that this anomalous scattering is a property of the chiral edge currents themselves, resolved in real space around defects of the lattice and active only in the dominant $\nu_{\textrm{off}}$ regime; it should not be conflated with the chiral inelastic photon-transport mechanism of~\cite{Peano2016} in which a probe field injected on the edge of a kagome cavity array emerges at a shifted frequency for any finite squeezing. Such current morphology is also seen on the outer edge of the kagome flake paralleling the bulk defect case (see details in~\cite{SM}).

\begin{figure}[!t]
  \centering
  \includegraphics[width=0.98\columnwidth]{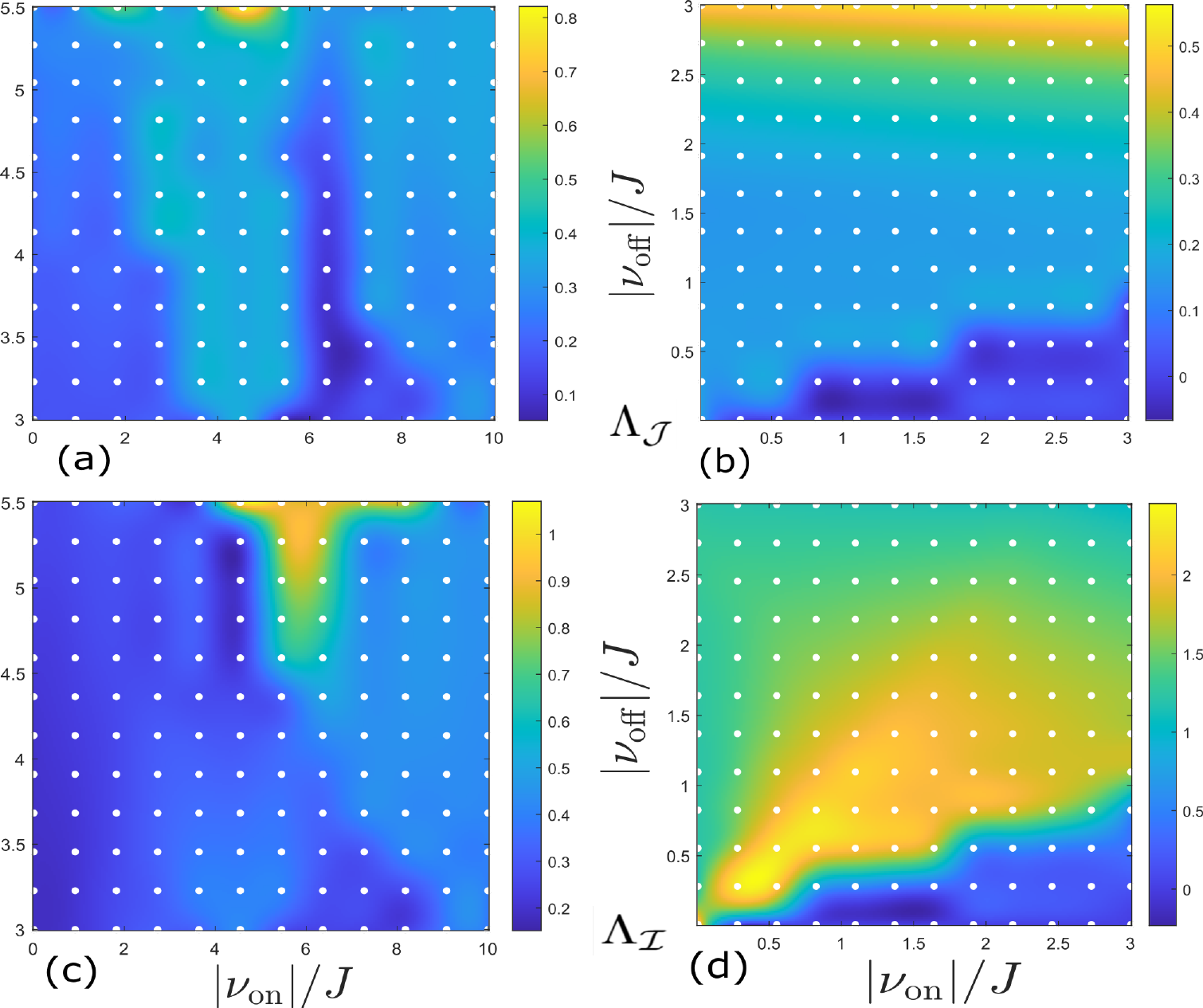}
  \caption{Clear signature of flux-engineering effects quantified via $\Lambda_{\cal J}$ and $\Lambda_{\cal I}$ on the plane spanned by a range of  $(\nu_{\non},\nu_{\noff})$ values compared between Case 1 and Case 3. The white dots are the exact numerical results while the colormap is derived via an interpolating function. For Case~1 both $\Lambda_{\cal J}$
  (a) and $\Lambda_{\cal I}$ (c), the leakage is suppressed throughout the plots; enhancement starts occurring only at unusually high values of BP coupling. Remarkably, Case~3 exhibits a sharp rise in anomalous scattering of PC around the line
  $\Lambda_{\cal I} \approx 1$ (d) a feature absent in $\Lambda_{\cal J}$ and the leakage increases monotonically (c). Here, as anticipated before, the confined regime is dominated by the relative magnitude of the on-site bosonic pairing coupling ($\nu_{\non}\gg\nu_{\noff}$)
 while strengthening of the off-site coupling leads to the anomalous regime ($\nu_{\noff}\gg\nu_{\non}$).}
  \label{fig:lambda-sweep}
\end{figure}

This distinct nature of the spatial pattern of currents motivates one to develop a numerical quantification of the confinement/leakage phenomenology. To analyze this crossover behavior, we define $\mathcal{J}_{\chiralcur}$ and  $\mathcal{I}_{\chiralcur}$ as the absolute SPC/PC flowing inside the 14-site loop around the
defect, and $\mathcal{J}_{\as}$ and $\mathcal{I}_{\as}$ as the absolute currents leaking into the neighboring sites outside the loop (a visual guide and mathematical detail is given in ~\cite{SM}).  The ratios
\begin{equation}
\Lambda_{\cal J}=\frac{\mathcal{J}_{\as}}{\mathcal{J}_{\chiralcur}},\qquad
\Lambda_{\cal I}=\frac{\mathcal{I}_{\as}}{\mathcal{I}_{\chiralcur}}
\label{eq:Lambda}
\end{equation}
measure leakage: small $\Lambda$ signals chiral confinement, large
$\Lambda$ the anomalous regime. Figure~\ref{fig:lambda-sweep} scans
both ratios on the BP-coupling plane. To highlight the differences we compare Case 1 (Figs.~\ref{fig:lambda-sweep} (a)-(c)) where the complex phase is only associated with the TB sector with that of Case 3 (Figs.~\ref{fig:lambda-sweep} (b)-(d)) that also includes strong TRS breaking on the BP sector. Significantly we find that in Case~3, the PC leakage ratio $\Lambda_{\cal I}$ develops a strong enhancement around $\Lambda_{\cal I} \approx 1$, which is absent in $\Lambda_{\cal J}$, where the increase is clearly monotonic and magnitude of anomalous scattering is also low. In Case~1, such anomalous behavior is absent and the confined nature of edge currents is distinctly revealed in the plots.

\parahead{Non-trivial bulk topology coexists with anomalous
leakage}%
A central conceptual conclusion of this Letter, made explicit by the
two-current decomposition, is that the enhanced leakage observed in
the anomalous regime is \emph{not} a loss of bulk topology. Both
regimes of Case~3 are bulk-gapped and carry integer para-unitary Chern
numbers, $(C_{1},C_{2},C_{3})=(+1,-1,0)$ in the anomalous regime and
$(+2,-1,+1)$ in the confined one---we provide details of the calculation of these numbers in \cite{SM} (Sec.~V). The two phases belong to distinct topological classes separated by bulk-gap
closings elsewhere in parameter space. The chirality of the edge
currents on a finite zigzag flake persists across both regimes and
tracks the sign of the lowest-band Chern number; what the BP sector
controls is the relative weight with which that chirality registers in
the two current operators, i.e.\ the ratio $\Lambda_{\cal I}$. This ratio
is a phase-sensitive observable: it varies smoothly with the
Hamiltonian phases $(\psi,\varphi)$---which do not close the bulk gap except along the codimension-one loci already characterized in~\cite{SM} (Sec.~V)---and with
$\nu_{\non}/\nu_{\noff}$ across the full parameter plane, including
regions of distinct bulk-Chern assignment, while the bulk gap stays
open.
Anomalous defect scattering is therefore an observable \emph{of} a chiral current carried by a bulk-gapped phase with non-trivial Chern indicators, not a breakdown of that phase---a distinction unavailable in
particle-conserving settings, where only a single current operator
exists.

\parahead{Experimental outlook}%
Two predictions discriminate our scenario from any
particle-conserving one: (a) chiral circulation of the pairing current
$\langle  \mathcal{I}\rangle$ along the edge of a finite zigzag-terminated
sample at vanishing hopping phase ($\psi=0$) but non-trivial pairing
phase, and (b) a qualitative reorganisation of the chiral current
pattern between the anomalous and confined regimes of
Case~3---reflected in both the spatial profile of $\langle
\mathcal{I}\rangle$ and the leakage ratio $\Lambda_{\mathcal{I}}$ defined in
Eq.~\eqref{eq:Lambda}---consistent with the Chern reorganisation
$(C_{1},C_{2},C_{3})=(+1,-1,0)\to(+2,-1,+1)$.
Both are accessible in a kagome circuit-QED array of parametrically
driven microwave resonators~\cite{Pocklington2023}---the bosonic,
resonator-based branch of circuit-QED, where each site hosts a
harmonic LC mode $a_{j,s}$ rather than a hard-core qubit---with $\psi$
set by SQUID flux and ($\nu_{\non},\nu_{\noff},\varphi$) implemented
as one- and two-mode microwave squeezing via parametric pumps at
$2\omega_{0}$. The primary, calibration-free signature is
$\Lambda_{\cal I}$, or equivalently its pointwise counterpart $\mathcal{R}=|\langle  \mathcal{I}\rangle/\langle  \mathcal{J}\rangle|$ at a
defect site, whose anomalous enhancement has no particle-conserving
analogue. Photonic lattices~\cite{Hafezi2013,Hafezi2019review} and
polariton arrays~\cite{Gulevich2017} provide complementary platforms;
details are in~\cite{SM}.

\parahead{Summary and conclusions}%
We have shown that bosonic pairing opens a second chiral current channel on a 2D kagome lattice, absent in any particle-conserving model on the same geometry and carrying an anomalous defect-scattering signature tunable through the phase structure of the Hamiltonian. The continuity equation assigns a direct microscopic meaning to this second channel as the pairing current $\langle \mathcal{I}\rangle$ itself, sourced by the off-site anomalous coherence $\langle a_{\vj}a_{\vj'}\rangle$ rather than by the on-bond single-particle coherence $\langle a_{\vj}^{\dagger} a_{\vj'}\rangle$, and visible in real space as a chiral edge circulation and as an anomalous scattering pattern around localized defects. The distinct response of the two currents to defects originates in the inequivalence of the single-particle and BdG pairing gaps in the anomalous regime, the microscopic mechanism being analyzed in detail in~\cite{SM} (Sec.~V).

The bulk Chern indicators are integer and stable within each regime and consistent with the observed chirality, although the standard fermionic bulk-boundary correspondence does not transfer literally to the bosonic BdG setting~\cite{SM}; the phenomenology we report is therefore primarily a statement about the \emph{currents}, and only indirectly about the invariant that labels their host phase. These results reposition bosonic pairing from a quantitative correction to a primary resource for the engineering of chiral transport, and identify a regime in which a bulk-gapped phase with non-trivial Chern indicators sustains a chiral current that is simultaneously robust to local defects and anomalously responsive to them---a coexistence that has no analogue in particle-conserving topological matter and that is, we argue, the generic fingerprint of pairing-enabled edge transport, for which the real-space currents, rather than the ribbon spectrum, are the appropriate diagnostic. Driven photonic lattices~\cite{Peano2016} and superconducting-circuit arrays~\cite{Pocklington2023}, where both phase structures can be engineered independently and the two channels can in principle be read out separately, provide natural experimental platforms to test this phenomenology.

\parahead{Acknowledgements}%
The authors acknowledge support from Spanish MICIN grants PID2021-122547NB-I00, EUR2024-15354 and the CAM Programa TEC-2024/COM-84 QUITEMAD-CM. This work has been financially supported by the Ministry for Digital Transformation and of Civil Service of the Spanish Government through the QUANTUM ENIA project call -- Quantum Spain project, and by the European Union through the Recovery, Transformation and Resilience Plan -- NextGenerationEU within the framework of the Digital Spain 2026 Agenda. We also acknowledge the ELLIS Unit Madrid. M.A.M.-D.\ has been partially supported by the U.S. Army Research Office through Grant No.\ W911NF-14-1-0103.

%% file: sm_body.tex
\section{Bogoliubov--de Gennes diagonalisation and numerical
methods}\label{sec:SM-I}

The quadratic bosonic Hamiltonian of the main text,
Eq.~(1), is Hermitian as an operator on the bosonic Fock
space, and the associated Bogoliubov--de Gennes (BdG) matrix
$D$ is a Hermitian $2m\!\times\!2m$ matrix on the doubled
Nambu space. What is non-standard in the bosonic case is its
\emph{diagonalisation}: the similarity transformation that
brings $D$ to canonical form must preserve the bosonic
commutation algebra and therefore cannot be an ordinary
unitary rotation. The correct framework is para-unitary
diagonalisation with respect to the paravector metric
$\Sigma_{z}=\operatorname{diag}(\mathbf{1}_{m},-\mathbf{1}_{m})$,
introduced by Colpa~\cite{SM:Colpa1978} and applied to
topological bosonic BdG systems by Shindou \emph{et
al.}~\cite{SM:Shindou2013}. We summarise the implementation used
in this work and record a specialised shortcut that exploits
the commutativity of the hopping and pairing blocks in our
kagome model.

\subsection{Para-unitary diagonalisation}\label{sec:SM-I-paraU}

We write the general quadratic bosonic Hamiltonian as
\begin{equation}
  H \;=\; \mathbf a^{\dagger} D\, \mathbf a,
  \qquad
  \mathbf a \;=\; (a_{1},\ldots,a_{m},a_{1}^{\dagger},\ldots,a_{m}^{\dagger})^{\intercal},
  \label{eq:SM-I-Hquad}
\end{equation}
with $D$ a Hermitian $2m\!\times\!2m$ matrix built out of the four
canonical blocks $(D_{1},D_{2},D_{3},D_{4})$ associated with
$a^\dagger a$, $a^\dagger a^\dagger$, $a a$, and $a a^\dagger$ terms.
The canonical commutation relations are encoded in the
paravector metric
\begin{equation}
  \Sigma_{z} \;=\; \operatorname{diag}(\mathbf{1}_{m},\, -\mathbf{1}_{m}),
  \qquad
  [\mathbf a_{r}, \mathbf a_{r'}^{\dagger}] \;=\; (\Sigma_{z})_{rr'}.
  \label{eq:SM-I-metric}
\end{equation}
A quasiparticle transformation $\mathbf c = T\mathbf a$, with
$\mathbf c = (\gamma_{1},\ldots,\gamma_{m},
\gamma_{1}^{\dagger},\ldots,\gamma_{m}^{\dagger})^{\intercal}$,
diagonalises $H$ iff
\begin{equation}
  (T^{\dagger})^{-1} D\, T^{-1}
  \;=\; E
  \;=\; \hbar\,\operatorname{diag}(\omega_{1},\ldots,\omega_{m},
                                    \omega_{1},\ldots,\omega_{m}),
  \label{eq:SM-I-diag}
\end{equation}
while preservation of the bosonic algebra requires the
para-unitarity condition
\begin{equation}
  T \,\Sigma_{z}\, T^{\dagger} \;=\; \Sigma_{z}.
  \label{eq:SM-I-paraunitary}
\end{equation}
A positive-definite $D$ is equivalent to the
\emph{dynamical stability} of the quadratic system.
Under this assumption the problem can be recast as an ordinary
Hermitian eigenvalue problem: writing $D = K^{\dagger}K$
(Cholesky), one diagonalises unitarily the Hermitian matrix
$K \Sigma_{z} K^{\dagger}$ and reconstructs the para-unitary $T$ from
its eigenvectors~\cite{SM:Colpa1978,SM:Shindou2013}. All matrices
encountered in this work are dynamically stable on the
parameter ranges used below, so this procedure applies.

\subsection{Faster route for the cases in question}\label{sec:SM-I-shortcut}

The Hamiltonian of the main text, Eq.~(1), produces a BdG matrix of
the block form
\begin{equation}
  D \;=\;
  \begin{pmatrix}
    A & B \\ B^{\dagger} & A
  \end{pmatrix},
  \label{eq:SM-I-blocks}
\end{equation}
with $A$ Hermitian and $B$ normal. For all three parameter
configurations (Cases 1, 2, 3) of Sec.~\ref{sec:SM-II} the hopping
and pairing blocks commute, $[A,B]=0$, because they are built from
the same lattice translations. In this regime a Cholesky
factorisation is not needed; $A$, $B$, $B^{\dagger}$, and
$\tilde D := A^{2}-B^{\dagger}B$ admit a common orthonormal
eigenbasis $\{\xi_{r}\}$, with $\tilde D\,\xi_{r}=\lambda_{r}^{2}\xi_{r}$.
The para-unitary eigenvectors of $D$ can be written in closed form,
\begin{equation}
  w_{r} \;=\; C_{r}
  \begin{pmatrix}
    (A+\lambda_{r}\mathbf{1})\,\xi_{r} \\[2pt]
    -B^{\dagger}\xi_{r}
  \end{pmatrix},
  \qquad
  w_{m+r} \;=\; C_{m+r}
  \begin{pmatrix}
    -B\,\xi_{r} \\[2pt]
    (A+\lambda_{r}\mathbf{1})\,\xi_{r}
  \end{pmatrix},
  \label{eq:SM-I-closed}
\end{equation}
with normalisation $u_{r}^{\dagger}u_{r}-v_{r}^{\dagger}v_{r}=+1$
(and $=-1$ for the $m+r$ branch), so that
$D\,w_{r}=+\lambda_{r}\,w_{r}$ and $D\,w_{m+r}=-\lambda_{r}\,w_{m+r}$.
Assembling $T^{-1}=(w_{1},\ldots,w_{2m})$ yields the
para-unitary diagonaliser of Eq.~\eqref{eq:SM-I-diag} without
invoking a Cholesky step. This shortcut is what the numerical
simulations of this work use; we have verified agreement with the
general Cholesky route to machine precision at representative
$\mathbf k$ points and finite-size configurations.

\subsection{Numerical protocol and validation}\label{sec:SM-I-num}

Real-space current maps in the main text and in
Sec.~\ref{sec:SM-IV} below are obtained by evaluating
Eqs.~(2) and (3) of the main text on eigenvectors of $D$
constructed through Eq.~\eqref{eq:SM-I-closed}.

For the momentum-space analysis (Sec.~\ref{sec:SM-VI}) we
translate the same BdG construction to the Bloch Hamiltonian
$H(\mathbf k)$ of the bulk lattice (periodic boundary
conditions in both primitive directions, i.e.\ the 2-torus),  diagonalise it with the para-unitary routine above on a
$51\!\times\!51$ Brillouin-zone grid, and compute the Chern numbers
of the three positive-frequency bands with the
Fukui--Hatsugai--Suzuki (FHS) link-variable algorithm~\cite{SM:FHS2005}
adapted to the para-unitary inner product,
$\langle u | v\rangle_{\Sigma}\equiv\langle u |\Sigma_{z}| v\rangle$
with $\Sigma_{z}=\operatorname{diag}(\mathbf{1}_{3},-\mathbf{1}_{3})$ in this case. Ribbon spectra are obtained on a zigzag-terminated slab of $L=18$
unit cells along one primitive direction with periodic boundary
conditions along the other. By \emph{zigzag termination} we mean
the canonical kagome convention: the slab is finite along the
primitive direction $\mathbf a_{2}=(1/2,\sqrt{3}/2)\,a$ with
$L=18$ unit cells, and periodic along
$\mathbf a_{1}=(1,0)\,a$, which retains $k_{x}$ as a good Bloch
momentum. Each of the two boundaries then exposes an alternating
$A$--$B$ row of kagome sites, while the $C$ sublattice is the one
whose inter-cell bonds ($C\!\to\!A$ and $C\!\to\!B$) are
truncated at the boundary. The alternative cut, parallel to
$\mathbf a_{1}+\mathbf a_{2}$, would produce an armchair edge and
is not used here. The triangular flake used for the
real-space current maps is \emph{not} employed to evaluate any
Chern number; its topological content is inferred from the bulk
invariants and ribbon bulk--boundary correspondence, as detailed
in Sec.~\ref{sec:SM-VI}. All three edges of the triangular flake
are zigzag-terminated in the same sense as the ribbon: each edge
is parallel to one of the primitive directions
$\mathbf a_{1}$, $\mathbf a_{2}$, $-(\mathbf a_{1}+\mathbf a_{2})$,
and the three $C_{3}$-equivalent corners terminate at single
isolated sites, which is the diagnostic fingerprint of three
concurrent zigzag cuts rather than armchair ones.
Integer convergence of the Chern
numbers has been checked by increasing the BZ grid from
$31\!\times\!31$ to $51\!\times\!51$ with no change to four decimal
places, and by verifying that the pairing-off limit reproduces the
tight-binding kagome Chern sequence $(0,-1,+1)$ at $\pi/2$
flux~\cite{SM:Peano2016,SM:Chaudhary2021}.

\section{The three phase configurations: Cases 1, 2, and
3}\label{sec:SM-II}

\begin{figure}[h!]
\centering
\includegraphics[width=0.95\textwidth]{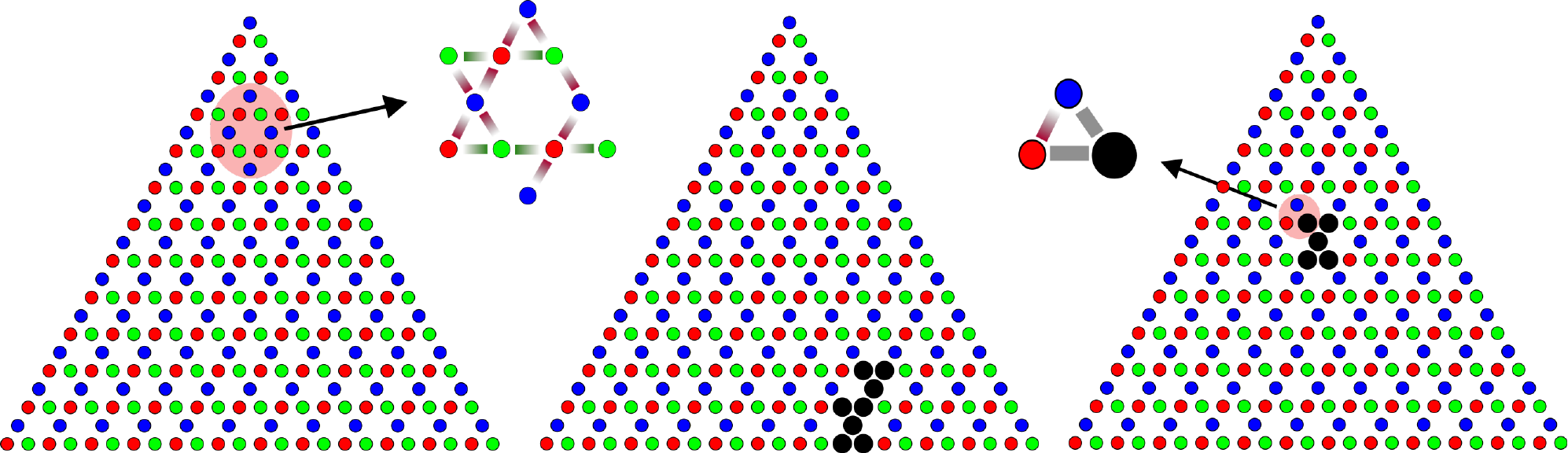}
\caption{The three triangular lattices from left to right used in the paper: a clean triangular flake with zigzag open boundary conditions, the same flake with edge defect sites and last one with bulk defect sites. If we zoom in, we see the three colours (R/G/B) of individual sites (A/B/C) denote the fact that they can have different phases associated with local on-site bosonic pairing terms $\frac{|\nu_{\non}}{2}|\bigl(e^{i\phi_{s}}a^{\dagger}_{\vj,s}a^{\dagger}_{\vj,s})$. Also the coloured bonds between individual sites (shadowed R/G/B) denote the presence of distinct phases on individual NN bonds arising from terms like $ - J\bigl(a^{\dagger}_{\vj,s}a_{\vj',s'}e^{-i\psi_{s,s'}}\big) - \frac{|\nu_{\noff}|}{2}\bigl(e^{-i\phi_{s,s'}}a^{\dagger}_{\vj,s}a^{\dagger}_{\vj',s'}\bigr)$. The defect sites are marked in black; the grey bonds connecting them to the normal sites are unaccessible at normal energies due to the tuning of the onsite energy of the defects to $\omega_{0}\approx 10^{4}$.}
\label{fig:SM-I}
\end{figure}

In the sections below,  we index lattice sites by $(\vj,s)$, where $\vj$ labels the Bravais unit cell and $s\in\{A,B,C\}$ the kagome sublattice. A bond is specified by a pair $(\vj,s),(\vj',s')$ with $\vj\neq\vj'$ or
$s\neq s'$ for nearest-neighbour (NN) links (see Fig. \ref{fig:SM-I}).   Hopping phases are denoted $\psi_{s,s'}$ and pairing phases $\phi_{s,s'}$ (off-site) or $\phi_{s}$ (on-site). The uniform flux quantum is $\alpha=2\pi/3$, so that a triangular plaquette with all three NN phases equal to $\alpha$ encloses a total flux of $2\pi$ and reduces to a single $\pi/2$-flux Haldane pattern once the gauge is redistributed on the
honeycomb-like bipartition of the kagome lattice, in direct analogy
with the tight-binding models of Refs.~\cite{SM:Peano2016,SM:Chaudhary2021}.\smallskip

\paragraph*{Case 1. --- Uniform hopping phase, real pairing.}
The Hamiltonian is
\begin{align}
H_{S}^{(1)} &= \frac{\hbar}{2}\!\left[\sum_{\vj,s}\omega_{0}\,a^{\dagger}_{\vj,s}a_{\vj,s}
- J\!\sum_{\langle(\vj,s),(\vj',s')\rangle}\!\bigl(a^{\dagger}_{\vj,s}a_{\vj',s'}e^{-i\psi_{s,s'}}
+\text{h.c.}\bigr)\right] \notag\\
&\quad
-\frac{\hbar}{2}\!\left[|\nu_{\non}|\!\sum_{\vj,s}\!\bigl(a^{\dagger}_{\vj,s}a^{\dagger}_{\vj,s}+\text{h.c.}\bigr)
+|\nu_{\noff}|\!\sum_{\langle(\vj,s),(\vj',s')\rangle}\!\bigl(a^{\dagger}_{\vj,s}a^{\dagger}_{\vj',s'}
+\text{h.c.}\bigr)\right],
\label{eq:SM-II-case1}
\end{align}
with $\psi_{AB}=\psi_{BC}=\psi_{CA}=\alpha=2\pi/3$ and $\phi_{s}=
\phi_{s,s'}=0$. Case~1 thus carries a uniform synthetic flux in the
hopping sector and only real (phase-free) bosonic pairing.\smallskip

\paragraph*{Case 2. --- Real hopping, spatially varying pairing phases.} Now the Hamiltonian reads
\begin{align}
H_{S}^{(2)} &= \frac{\hbar}{2}\!\left[\sum_{\vj,s}\omega_{0}\,a^{\dagger}_{\vj,s}a_{\vj,s}
- J\!\sum_{\langle(\vj,s),(\vj',s')\rangle}\!\bigl(a^{\dagger}_{\vj,s}a_{\vj',s'}+\text{h.c.}\bigr)\right] \notag\\
&\quad
-\frac{\hbar}{2}\!\left[|\nu_{\non}|\!\sum_{\vj,s}\!\bigl(e^{i\phi_{s}}a^{\dagger}_{\vj,s}a^{\dagger}_{\vj,s}+\text{h.c.}\bigr)
+|\nu_{\noff}|\!\sum_{\langle(\vj,s),(\vj',s')\rangle}\!\bigl(e^{-i\phi_{s,s'}}a^{\dagger}_{\vj,s}a^{\dagger}_{\vj',s'}+\text{h.c.}\bigr)\right],
\label{eq:SM-II-case2}
\end{align}
with $\psi_{s,s'}=0$ and the sublattice-dependent pairing phases
$\phi_{A}=0$, $\phi_{B}=\alpha$, $\phi_{C}=2\alpha$,
$\phi_{AB}=0$, $\phi_{BC}=\alpha$, $\phi_{CA}=2\alpha$. This is the
minimal model in which the only time reversal symmetry-breaking term is carried by the
BP sector; it isolates the pairing-induced topology. \smallskip

\paragraph*{Case 3. --- Both sectors dressed with phases.} In this general case
\begin{align}
H_{S}^{(3)} &= \frac{\hbar}{2}\!\left[\sum_{\vj,s}\omega_{0}\,a^{\dagger}_{\vj,s}a_{\vj,s}
- J\!\sum_{\langle(\vj,s),(\vj',s')\rangle}\!\bigl(a^{\dagger}_{\vj,s}a_{\vj',s'}e^{-i\psi_{s,s'}}+\text{h.c.}\bigr)\right] \notag\\
&\quad
-\frac{\hbar}{2}\!\left[|\nu_{\non}|\!\sum_{\vj,s}\!\bigl(e^{i\phi_{s}}a^{\dagger}_{\vj,s}a^{\dagger}_{\vj,s}+\text{h.c.}\bigr)
+|\nu_{\noff}|\!\sum_{\langle(\vj,s),(\vj',s')\rangle}\!\bigl(e^{-i\phi_{s,s'}}a^{\dagger}_{\vj,s}a^{\dagger}_{\vj',s'}+\text{h.c.}\bigr)\right],
\label{eq:SM-II-case3}
\end{align}
with $\psi_{AB}=\psi_{BC}=\psi_{CA}=\alpha=2\pi/3$ and the same
pairing phases as in Case~2. Within Case~3 we distinguish two
sub-regimes by the ratio of on-site to off-site pairing amplitudes:

\begin{itemize}
\item Case~3a: $\nu_{\non}\ll\nu_{\noff}$, parameter point
$(\nu_{\non},\nu_{\noff})=(0.01J,3J)$.
\item Case~3b: $\nu_{\non}\gg\nu_{\noff}$, parameter point $(3J,0.01J)$.
\end{itemize}
The first hosts the anomalous-scattering regime, the second the confined-transport regime with $|C_{1}|=2$ (see Sec.~\ref{sec:SM-VI}).\\

\paragraph*{Parameter scales used throughout.}
Unless stated otherwise we fix $\omega_{0}=50$ and $J=0.02\,\omega_{0}$;
defect sites are imposed by setting $\omega_{\mathcal S}=10^{4}\omega_{0}$
on a connected cluster $\{\mathcal S\}$ either on the edge or in the bulk. Representative BP couplings per Case are tabulated in Table~\ref{tab:chern-table} and are used identically in the current maps of Sec.~\ref{sec:SM-IV} and in the topological analysis of Sec.~\ref{sec:SM-VI}.

\section{Derivation of the two current operators from the continuity
equation}\label{sec:SM-III}

Setting $\hbar=1$ we split the general Hamiltonian of Case~3 [Eq.~\eqref{eq:SM-II-case3}] into an on-site piece
$H_{0}=\omega_{0}\sum_{\vj,s}a^{\dagger}_{\vj,s}a_{\vj,s}$ and two
bond operators,
\begin{align}
V_{\TB}&=-J\sum_{\langle(\vj,s),(\vj',s')\rangle}\!\bigl(a^{\dagger}_{\vj,s}a_{\vj',s'}\,e^{-i\psi_{s,s'}}
+\text{h.c.}\bigr),
\label{eq:SM-III-VTB}\\
V_{\BP}&=-\tfrac{1}{2}\Bigl[|\nu_{\non}|\sum_{\vj,s}\bigl(e^{i\phi_{s}}a^{\dagger}_{\vj,s}a^{\dagger}_{\vj,s}+\text{h.c.}\bigr)
+|\nu_{\noff}|\!\sum_{\langle(\vj,s),(\vj',s')\rangle}\!\bigl(e^{-i\phi_{s,s'}}a^{\dagger}_{\vj,s}a^{\dagger}_{\vj',s'}+\text{h.c.}\bigr)\Bigr].
\label{eq:SM-III-VBP}
\end{align}
A lattice continuity equation is obtained from the Heisenberg
equation of motion for the on-site population,
\begin{equation}
\frac{d}{dt}\langle a^{\dagger}_{\vj,s}a_{\vj,s}\rangle
=i\,\bigl\langle[H_{S},a^{\dagger}_{\vj,s}a_{\vj,s}]\bigr\rangle.
\label{eq:SM-III-continuity}
\end{equation}
Since $[H_{0},a^{\dagger}_{\vj,s}a_{\vj,s}]=0$, only the bond
operators contribute in Eq.~\eqref{eq:SM-III-continuity}.\smallskip

\paragraph*{Single-particle (hopping) contribution.}
Using the canonical commutators
$[a^{\dagger}_{\vj,s}a_{\vj',s'},a^{\dagger}_{\vl,r}a_{\vl,r}]
=a^{\dagger}_{\vj,s}a_{\vj',s'}(\delta_{(\vj,s),(\vl,r)}
- \delta_{(\vj',s'),(\vl,r)})$, we obtain
\begin{equation}
i[V_{\TB},a^{\dagger}_{\vj,s}a_{\vj,s}]
=-\!\!\sum_{(\vj',s')\neq(\vj,s)}\!\!\mathcal{J}_{(\vj,s)\to(\vj',s')},
\label{eq:SM-III-tbcomm}
\end{equation}
where the summation is over all $(\vj',s')$ which are only NN of $(\vj,s)$ and the current operator is identified as
\begin{equation}
\boxed{\;\mathcal{J}_{(\vj,s)\to(\vj',s')}
=iJ\!\left[a^{\dagger}_{\vj,s}a_{\vj',s'}e^{i\psi_{s,s'}}
-a^{\dagger}_{\vj',s'}a_{\vj,s}e^{-i\psi_{s,s'}}\right]\bigl(1-\delta_{\vj\vj'}\delta_{ss'}\bigr).\;}
\label{eq:SM-III-JSPC}
\end{equation}
The factor $(1-\delta_{\vj\vj'}\delta_{ss'})$ eliminates
the trivial on-site contribution.
Equation~\eqref{eq:SM-III-JSPC} is the single-particle current
(SPC) of Eq.~(2) in the main text.\smallskip

\paragraph*{Pairing contribution.}
Let us partition the bosonic pairing potential into the on-site and off-site blocks, ie., $V_{\BP}=V^{\non}_{\BP}+V^{\noff}_{\BP}$. The on-site pairing block, via same-site commutators like  $[a^{\dagger}a^{\dagger},a^{\dagger}a]=
-2\,a^{\dagger}a^{\dagger}$, leads to a local sink/source term:
\begin{equation}
i[V^{\non}_{\BP},a^{\dagger}_{\vj,s}a_{\vj,s}]
=i\,|\nu_{\non}|\!\left[a^{\dagger}_{\vj,s}a^{\dagger}_{\vj,s}e^{i\phi_{s}}
-a_{\vj,s}a_{\vj,s}e^{-i\phi_{s}}\right] \mbox{  }\equiv \mathcal{S}_{(\vj,s)}
\label{eq:SM-III-bponcomm}
\end{equation}

More importantly, the off-site bosonic pairing
term contributes
\begin{equation}
i[V_{\BP}^{\noff},a^{\dagger}_{\vj,s}a_{\vj,s}]
=-\!\!\sum_{(\vj',s')\neq(\vj,s)}\!\!\mathcal{I}_{(\vj,s)\to(\vj',s')},
\label{eq:SM-III-bpcomm}
\end{equation}
where again the summation is over all $(\vj',s')$ which are only NN of $(\vj,s)$ and the current opeartor is identified as
\begin{equation}
\boxed{\;\mathcal{I}_{(\vj,s)\to(\vj',s')}
=i\,\frac{|\nu_{\noff}|}{2}\!\left[a^{\dagger}_{\vj,s}a^{\dagger}_{\vj',s'}e^{i\phi_{s,s'}}
-a_{\vj',s'}a_{\vj,s}e^{-i\phi_{s,s'}}\right]\bigl(1-\delta_{\vj\vj'}\delta_{ss'}\bigr).\;}
\label{eq:SM-III-IPC}
\end{equation}
Equation~\eqref{eq:SM-III-IPC} is the pairing current (PC) of
Eq.~(3) in the main text. Unlike $\mathcal{J}$, the pairing
current $\mathcal{I}$ does not conserve the total particle
number $\hat N=\sum_{\vj,s}a^{\dagger}_{\vj,s}a_{\vj,s}$:
being bilinear in two creation (or two annihilation) operators, it
changes $\hat N$ by $\pm 2$ each time it acts. Accordingly,
$\mathcal{I}$ cannot be written as the divergence of a
$U(1)$ Noether current---the pairing term $V_{\BP}^{\noff}$
explicitly breaks the global particle-number symmetry---and in
particular $\sum_{(\vj,s)\neq(\vj',s')}\mathcal{I}_{(\vj,s)\to(\vj',s')}\neq 0$.
It is nonetheless a legitimate local lattice current in the BdG
formalism: it is the unique bond-resolved operator whose
divergence reproduces the pairing contribution to
$d\langle a^{\dagger}_{\vj,s}a_{\vj,s}\rangle/dt$ via Heisenberg's
equation, i.e.\ it tracks the local rate at which the pairing term
injects or removes particle pairs through the bond
$(\vj,s)\!\leftrightarrow\!(\vj',s')$. The total energy
$V_{\TB}+V_{\BP}$ and the fermion-like parity $(-1)^{\hat N}$
remain conserved, as is standard for any number-non-conserving BdG
Hamiltonian.\smallskip

\paragraph*{Global consistency.}
Combining Eqs.~\eqref{eq:SM-III-continuity},
\eqref{eq:SM-III-tbcomm}, \eqref{eq:SM-III-bponcomm}, \eqref{eq:SM-III-bpcomm}, the local
continuity equation reads
\begin{equation}
\frac{d}{dt}\langle a^{\dagger}_{\vj,s}a_{\vj,s}\rangle
+\!\!\sum_{(\vj',s')\neq(\vj,s)}\!\!\bigl(\langle\mathcal{J}_{(\vj,s)\to(\vj',s')}\rangle
+\langle\mathcal{I}_{(\vj,s)\to(\vj',s')}\rangle\bigr)=\mathcal{S}_{(\vj,s)}
\label{eq:SM-III-balance}
\end{equation}
which is equivalent to the total energy-conservation identity
$d\langle V_{\TB}+V_{\BP}\rangle/dt
=-\omega_{0}\sum_{\vj,s}d\langle a^{\dagger}_{\vj,s}a_{\vj,s}\rangle/dt$
as can be verified directly. The steady-state maps displayed in
the main text and in Sec.~\ref{sec:SM-IV} are obtained by evaluating
$\langle\mathcal{J}\rangle$ and $\langle\mathcal{I}\rangle$
on individual eigenvectors of the para-unitary BdG matrix, summed
over all nearest-neighbour bonds $(\vj,s)\to(\vj',s')$ of a triangular
flake with open boundary conditions.

It is trivial to derive the equivalent expressions for cases~1 and 2  by taking into account the proper phase factors in each sector.

\section{Edge and bulk transport: complementary numerical
evidence}\label{sec:SM-IV}

This section collects the supporting numerical evidence that
complements the three figures of the main text. The four subsections
cover, in order: edge currents in Cases~1 and~2 in the clean flake
(no defect) corresponding to Fig.~2 of main text;  an isolated eigenmode of Case~1 that carries chiral currents simultaneously along edge of the lattice and bulk defects; the geometric construction behind the ratios $\Lambda_{\cal J}$ and
$\Lambda_{\cal I}$ of Eq.~(4) of the main text; and  complementary
edge-defect current maps the confined vs anomalous regime in Case~3.  Related Chern number calculation gien in Table~\ref{tab:chern-table}.

\subsection{Edge current without defects corresponding to Fig.  2 in the main text}
\label{sec:SM-IVA}

\begin{figure}[h!]
\centering
\includegraphics[width=0.80\textwidth]{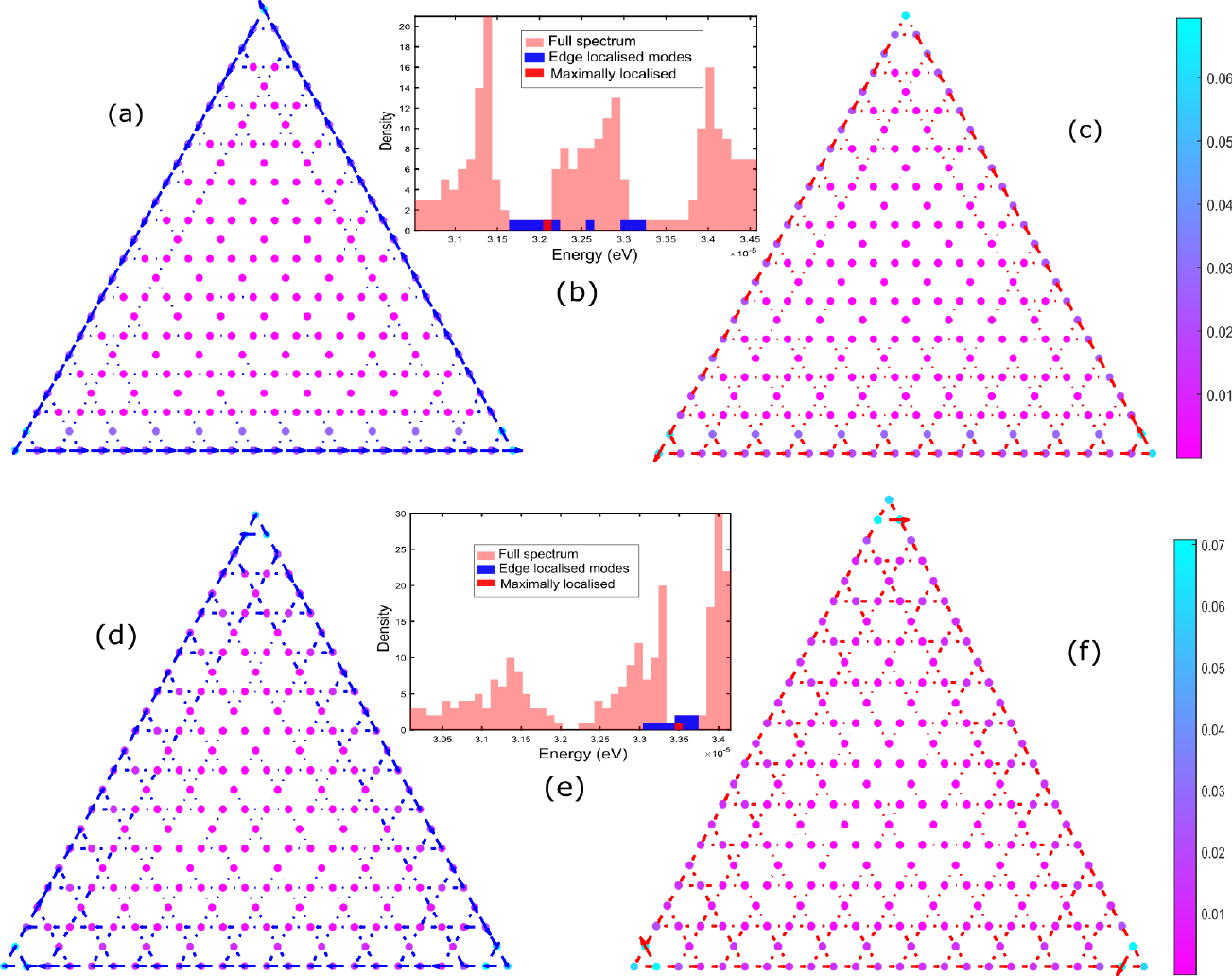}
\caption{SPC, eigenspectrum and PC for Case~1 are shown in panels (a),(b) and (c) respectively. Parameter values exactly as panels (a),(b),(e) and (f) of Fig.~2 in main text.  The corresponding SPC, eigenspectrum and PC for Case~2 are shown in panels (d),(e) and (f) respectively. Parameter values exactly as panels (c),(d),(g) and (h) of Fig.~2 in main text. The blue/red arrows indicate SPC/PC always. The colormap on the lattice gridpoints indicate the absolute squared value of the LEM wavefunction. Note the LEM-s sitting in the gap of the eigenspectrum.}
\label{fig:SM-IV-nodefect}
\end{figure}

Figure~\ref{fig:SM-IV-nodefect} shows the steady-state spatial distribution of SPC and PC for Case~1 and Case~2 on a clean triangular flake, i.e.  \emph{without} any defect.  The edge supports a chiral current
in both channels, establishing that the pairing-induced topological
edge mode is intrinsic to the clean lattice: no on-site detuning is
required to activate it.  The spatial modulation visible along the
edge in the current patterns of Case~2 already reflects  the onset of anomalous scattering for the edge modes due to the presence of phases in the BP sector.

\subsection{Inter-loop eigenmode in Case~1}
\label{sec:SM-IVB}

\begin{figure}[t]
\centering
\includegraphics[width=0.84\textwidth]{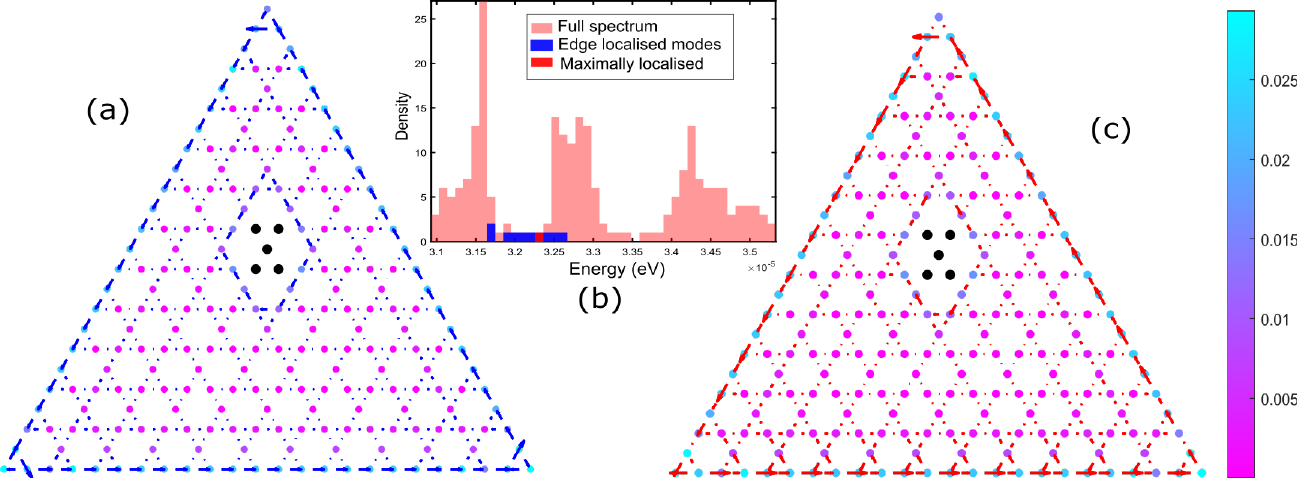}
\caption{Case~1, inter-loop eigenmode.
Single-particle (a) and pairing (c) current maps of the
isolated BdG eigenmode that sustains chiral edge current on the outer edge and the bulk
defect loop in the Case~1. Parameters: Parameters: $\omega_{0}=50$, $J=0.02\,\omega_{0}$,
  $\nu_{\non}=0.15J$, $\nu_{\noff}=0.30J$. Defect sites' chemical potential detuned to $10^{4} \omega_{0}$.}
\label{fig:SM-IV-cross}
\end{figure}

In a narrow parameter window of Case~1 we find an isolated eigenmode
of the BdG spectrum that carries chiral currents simultaneously along edge of the lattice and bulk defects.  Figure~\ref{fig:SM-IV-cross} displays the
SPC and PC maps of this mode.  Pannel (b) shows the eigenspectrum of the clean Hamiltonian without defect; notice the maximally localised edge mode sitting in a low-degeneracy gap in the spectrum.  The SPC shows two co-rotating chiral paths, one on the outer edge and one around the bulk defect; PC exhibits the same double-loop structure although anomalous scattering starts appearing at both edges.

\subsection{Geometric scheme for $\Lambda_{\cal J}$ and $\Lambda_{\cal I}$}
\label{sec:SM-IVC}

\begin{figure}[h!]
\centering
\includegraphics[width=0.45\textwidth]{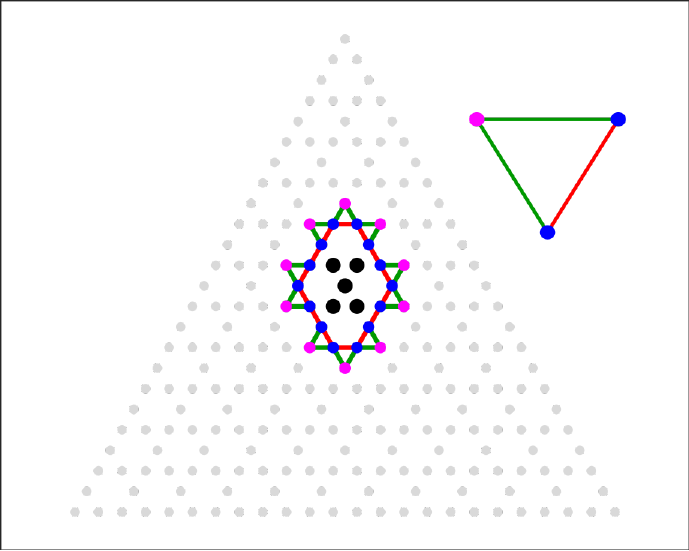}
\caption{ Geometric scheme for the leakage ratios
$\Lambda_{\cal J}$ and $\Lambda_{\cal I}$.}
\label{fig:SM-IV-cartoon}
\end{figure}

 Figure~\ref{fig:SM-IV-cartoon} gives the geometric construction used
in the main text to define the leakage ratios. On the flake with a
connected defect cluster $\{\mathcal S\}$, the $14$ innermost Kagome
sites adjacent to $\{\mathcal S\}$ form a closed loop $\mathcal L$
(blue sites conneted by red bonds). The sum of absolute values of the currents along the bonds of $\mathcal L$ defines $\mathcal J_{\chiralcur}\!=\!\sum_{\mathcal L}|\langle\mathcal{J}\rangle|$ and $\mathcal I_{\chiralcur}\!=\!\sum_{\mathcal L}|\langle\mathcal{I}\rangle|$. The nearest-neighbour bonds of $\mathcal L$ that leave the loop (magenta sites and connected to loop $\mathcal L$ by green bonds) define the leakage $\mathcal J_{\as}$ and $\mathcal I_{\as}$, so that
\begin{equation}
\Lambda_{\cal J}\equiv \frac{\mathcal J_{\as}}{\mathcal J_{\chiralcur}},
\qquad
\Lambda_{\cal I}\equiv \frac{\mathcal I_{\as}}{\mathcal I_{\chiralcur}},
\label{eq:SM-IV-lambda}
\end{equation}
coinciding with Eq.~(4) of the main text.  Both ratios are dimensionless and independent of the overall normalisation of the BdG eigenvector.

\subsection{Edge defect in Case~3: confined vs.\ anomalous regimes}
\label{sec:SM-IVD}

\begin{figure}[h]
\centering
\includegraphics[width=0.85\textwidth]{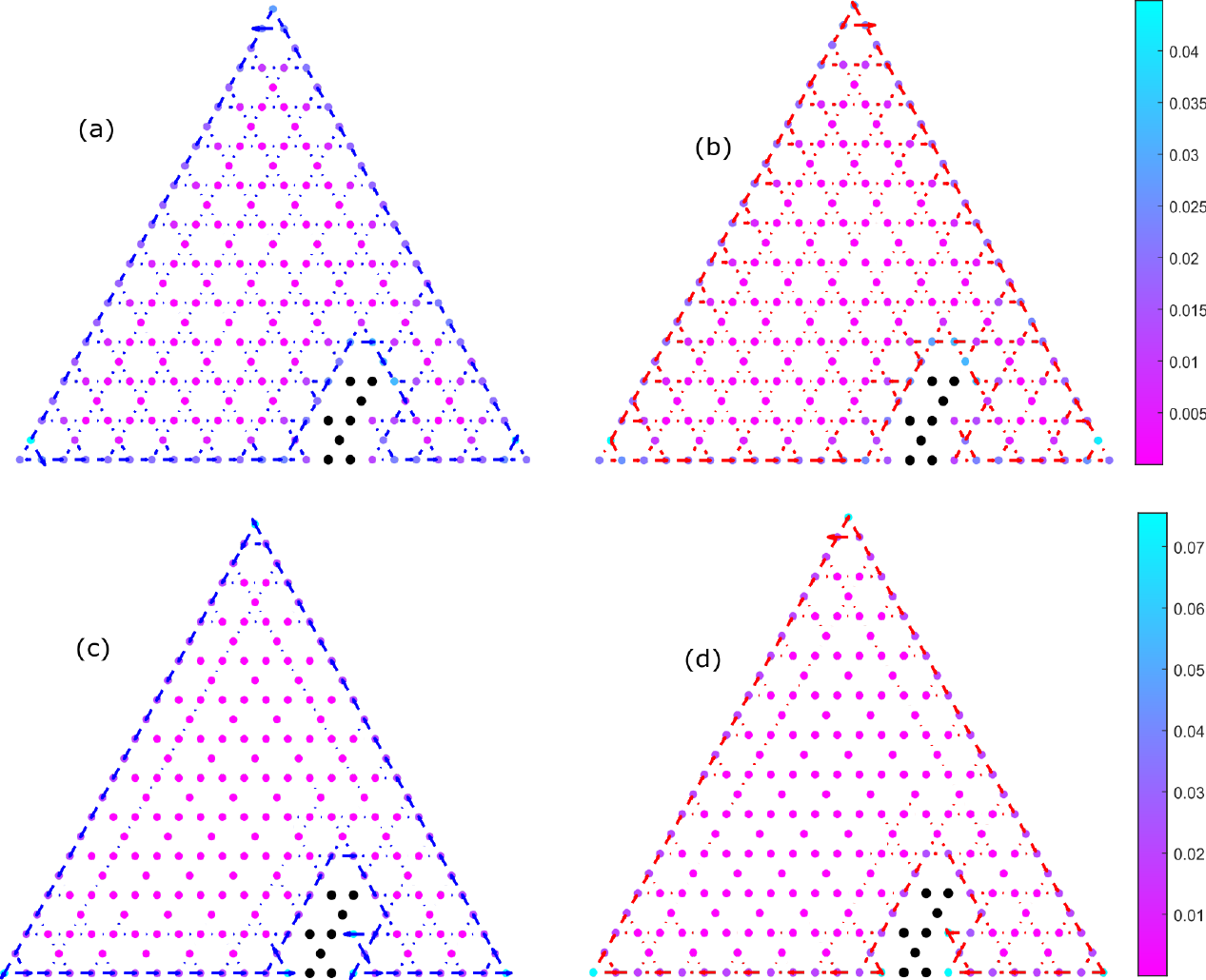}
\caption{Case~3 with an edge defect. Panels~(a)--(b): anomalous
regime, Case~3a $(\nu_{\non},\nu_{\noff})=(0.01J,3J)$, SPC and PC.
Panels~(c)--(d): confined regime, Case~3b
$(\nu_{\non},\nu_{\noff})=(3J,0.01J)$, SPC and PC. Both regimes are
bulk-gapped and carry integer para-unitary Chern numbers
(Table~\ref{tab:chern-table}): $(C_{1},C_{2},C_{3})=(+1,-1,0)$ in
Case~3a and $(+2,-1,+1)$ in Case~3b. The qualitative difference
reflects two distinct bulk topologies: the confined regime carries
$|C_{1}|\!=\!2$, the largest first-Chern in our parameter set, and
the tight chiral circulation seen in~(c)--(d) is the
non-perturbative real-space imprint of this enhancement; the
anomalous regime has $|C_{1}|\!=\!1$, and the strong PC leakage
visible in~(a)--(b) is not a loss of topological protection.}
\label{fig:SM-IV-edgeCase3}
\end{figure}

Figure~\ref{fig:SM-IV-edgeCase3} isolates the two regimes of Case~3
for a connected edge-defect cluster $\{\mathcal S\}$. In the
anomalous regime [$\nu_{\noff}\gg\nu_{\non}$, panels (a)--(b)] the
PC channel shows strong leakage into the bulk around the defect,
consistent with the $\Lambda_{\cal I}$ enhancement for a bulk defect of the main text. In
the confined regime [$\nu_{\non}\gg\nu_{\noff}$, panels (c)--(d)] the
currents remain sharply localised on the edge, in agreement with the topological protection established in Sec.~\ref{sec:SM-VI} via a nonzero Chern number for the lowest band $|C_{1}|=2$.

We emphasise that both regimes are topologically non-trivial. The
para-unitary Chern sequences reported in
Table~\ref{tab:chern-table} are integer to four decimal places on
the $51\times 51$ BZ grid in both regimes---$(C_{1},C_{2},C_{3})
=(+1,-1,0)$ in the anomalous regime (Case~3a) and $(+2,-1,+1)$ in
the confined regime (Case~3b). Accordingly, the enhanced leakage
seen in panels~(a)--(b) is \emph{not} a breakdown of topological
protection: the chiral edge channel of $C_{1}=+1$ is still present,
supported by the ribbon bulk--boundary correspondence of
Sec.~\ref{sec:SM-VI}.  The two regimes are thus physically distinguished not by the presence or absence of topology, but by the Chern multiplicity
$|C_{1}|\!=\!1$ vs.\ $|C_{1}|\!=\!2$ and by the relative weight
with which the surviving chiral channel reveals itself in the two
current operators---the observable precisely captured by the ratio
$\Lambda_{\cal I}$.

\subsection{Visualisation of confined mode vs leakage via anomalous scattering: a wavefunction perspective}

\begin{figure}[h]
\centering
\includegraphics[width=0.85\textwidth]{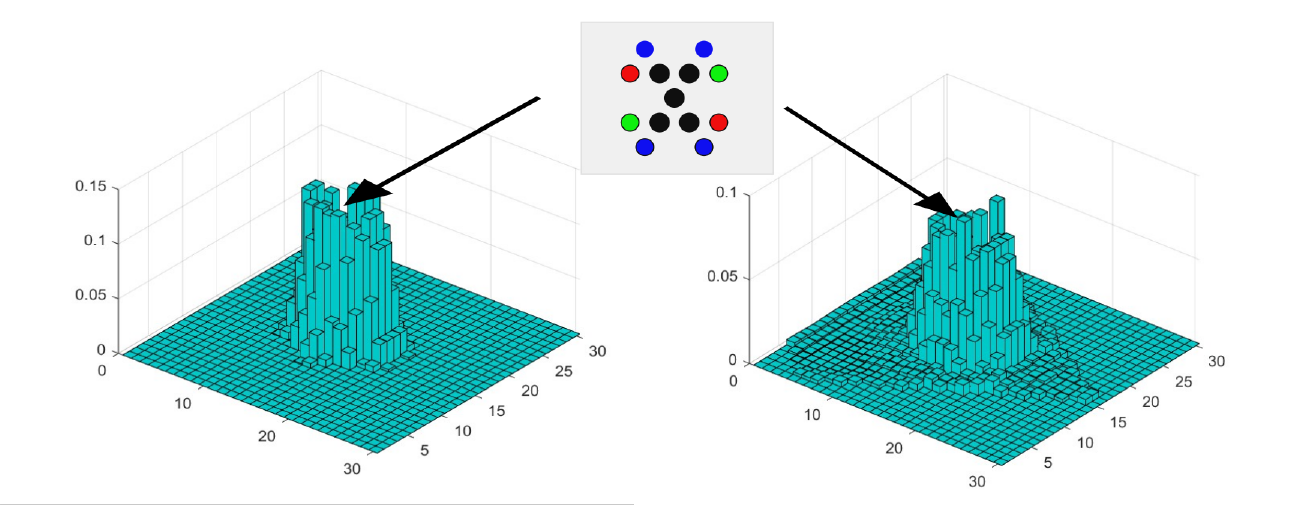}
\caption{Left panel shows the corresponding probability density (its magintude scaled to the height of the individual bars at each lattice site) map of the eigenmode that produces the confined chiral PC pattern around the bulk defect sites in Fig.3(b) of the main text. Right panel reveals that the eigenmode is considerably delocalised around the bulk defect sites signalling leakage of PC away from the innermost defect loop as evidenced in Fig.3(d) of main text (the 3D nature of the cityscape maps hide the exact location of the defect sites; hence the inset directs the reader's attention to the location of the defects in the lattice on which the probability map stands).}
\label{fig:SM-VI-wavefunctionCase3}
\end{figure}

Below we demonstrate the confined nature of the current (PC) corresponding to Fig. 3(b) and the anomalous scattering related to Fig, 3(d) of the main text via the following visualisation in Fig.~\ref{fig:SM-VI-wavefunctionCase3}: a cityscape view of the  probability density of the respective eigenmodes that create the distinct spatial features depending on the relative strength of the bosonic pairing couplings. The fine resolution in the height of the bars clearly demonstrate that in the confined regime, the wavefunction itself is also tightly localised around the bulk defect. While in the anomalous scattering regime, the wavefunction spreads considerably away from the defect sites as proved quantitatively via the measurement of leakage ratio-s  $\Lambda_{\cal J}$ and $\Lambda_{\cal I}$.

\section{Topological certification: Chern numbers and ribbon
spectra}\label{sec:SM-VI}

We certify the topological character of the edge channels reported
in the main text by computing the para-unitary Chern numbers of the
three positive-frequency bands and verifying the bulk--boundary
correspondence on a finite ribbon. The methodology follows
Fukui--Hatsugai--Suzuki (FHS)~\cite{SM:FHS2005} adapted to the bosonic
BdG inner product, as reviewed in Sec.~\ref{sec:SM-I-num} and
established by Shindou \emph{et al.}~\cite{SM:Shindou2013}.

\paragraph*{Geometries used in this work.}
Three distinct geometries appear in this Supplemental Material and
it is important to keep them conceptually separate:
\begin{itemize}[leftmargin=*,itemsep=3pt]
\item \emph{Bulk (2-torus).} Kagome lattice with periodic boundary
      conditions in \emph{both} primitive directions, i.e.\ the
      infinite lattice parametrised by quasimomentum
      $\vk\in\mathrm{BZ}$. This is the geometry on which the
      para-unitary Chern numbers $C_{n}$ of
      Eq.~\eqref{eq:SM-VI-FHS} are defined and computed.
\item \emph{Ribbon.} Zigzag-terminated slab of $L=18$ unit cells
      along one primitive direction, with periodic boundary
      conditions along the other. One direction retains a good
      Bloch momentum $k_{\parallel}$; the other is open. This
      geometry supports a well-defined edge spectrum and is the
      natural setting in which to verify bulk--boundary
      correspondence.
\item \emph{Triangular flake.} Finite kagome cluster with
      \emph{open} boundary conditions on all three sides (no
      translation symmetry in either direction). All real-space
      current maps of Figs.~1-3 of the main text and of
      Sec.~\ref{sec:SM-IV} are obtained on this geometry.
\end{itemize}
The Chern numbers reported below are computed exclusively on the
bulk 2-torus; no Chern number is computed directly on the
triangular flake. The topological identification of the edge currents observed on the flake rests on
three mutually reinforcing pieces of evidence:

(i) integer para-unitary Chern numbers in the bulk, evaluated on the same
periodic lattice,

(ii) where applicable, the standard fermionic bulk--boundary heuristic
[Eq.~\eqref{eq:SM-VI-BBC}] applied to the zigzag ribbon,

(iii) the direct non-perturbative observation of chiral circulation on
finite flakes (main text, Figs.~1--3), which is what the
experiment ultimately probes.

The bulk inputs to step (i) are robust: (a) the $C_{n}$ are integers to four decimal places on a
$51\times 51$ BZ grid and stable against doubling the resolution,
and (b) the pairing-off limit reproduces the tight-binding Chern
sequence $(0,-1,+1)$~\cite{SM:Peano2016,SM:Chaudhary2021}. The ribbon
step (ii) is verified at $L=18$ for Cases~1, 2 and 3a, but not at
the resolutions accessible to us for Case~3b, where three effects act
together: the gap $G_{2|3}$ is narrow, the upper band is quasi-flat,
and the bosonic BdG bulk--boundary correspondence is not literal in
the presence of non-topological kagome boundary states~\cite{SM:NakadaFujita1996,SM:Sekh2026,SM:Xia2020,SM:Gulevich2017}.
A proper treatment is given by the finite-size scaling analysis of
Sec.~\ref{sec:SM-VI-scaling}; this does not undermine the
topological identification, which is anchored in (i) and (iii)
rather than in a literal ribbon spectral-flow argument.\smallskip

\noindent \paragraph*{Chern numbers and edge-mode counts.}
The Chern number of band $n$ is
\begin{equation}
C_{n}=\frac{1}{2\pi}\!\sum_{\vk}\!F^{(n)}(\vk),
\qquad
F^{(n)}(\vk)=\operatorname{Arg}\!\bigl[U^{(n)}_{x}(\vk)U^{(n)}_{y}(\vk+\mathbf d_{x})\,U^{(n)*}_{x}(\vk+\mathbf d_{y})\,U^{(n)*}_{y}(\vk)\bigr],
\label{eq:SM-VI-FHS}
\end{equation}
with link variables
$U^{(n)}_{\mu}(\vk)
=\langle u_{n}(\vk)|\Sigma_{z}|u_{n}(\vk+\mathbf d_{\mu})\rangle
/|\langle u_{n}(\vk)|\Sigma_{z}|u_{n}(\vk+\mathbf d_{\mu})\rangle|$
and  $ \mathbf{d}_{\mu}$ is the basis vector in the direction $\mu$ in the recriprocal latrice.
The four representative parameter points tabulated in
Table~\ref{tab:chern-table} give the Chern sequences
$(+1,0,+1)$ (Case~1), $(+1,0,0)$ (Case~2), $(+1,-1,0)$ (Case~3a)
and $(+2,-1,+1)$ (Case~3b). All values are integers to four decimal
places on a $51\times 51$ BZ grid and stable against doubling the
resolution.\smallskip

\begin{table}[h!]
  \centering
  \begin{tabular}{lcc}
    \hline\hline
    Case & $(\nu_{\non},\nu_{\noff})$ & Chern $(C_{1},C_{2},C_{3})$ \\
    \hline
    1   & $(2J,\,4J)$         & $(+1, 0, +1)$ \\
    2   & $(0.1J,\,3J)$       & $(+1, 0, 0)$  \\
    3a  & $(0.01J,\,3J)$      & $(+1,-1, 0)$  \\
    3b  & $(3J,\,0.01J)$      & $(+2,-1,+1)$  \\
    \hline\hline
  \end{tabular}
  \caption{Integer para-unitary Chern numbers of the bulk bands for
  the four representative parameter configurations, computed with
  the Fukui--Hatsugai--Suzuki prescription on a $51\times 51$ BZ
  grid. The values are stable under modest variations of the
  pairing parameters; see Sec.~\ref{sec:SM-VI-scaling} for a
  discussion of how these bulk invariants relate to the ribbon
  spectra of Fig.~\ref{fig:chern-ribbon}. The parameter points used in
  Fig.~1 of the main text (Case~1, $(\nu_{\non},\nu_{\noff})=(3J,0.1J)$),
  Fig.~2 of the main text (Cases~1 and~2, $(0.1J,3J)$) and the inter-loop
  eigenmode of Fig.~\ref{fig:SM-IV-cross} (Case~1, $(0.15J,0.30J)$)
  all lie within the same topological phase as the canonical points
  tabulated here and therefore share the corresponding Chern triplet,
  by the parameter-space stability documented in
  Sec.~\ref{sec:SM-VI-scaling}.}
  \label{tab:chern-table}
\end{table}

\begin{figure}[h!]
  \centering
  \includegraphics[width=\textwidth]{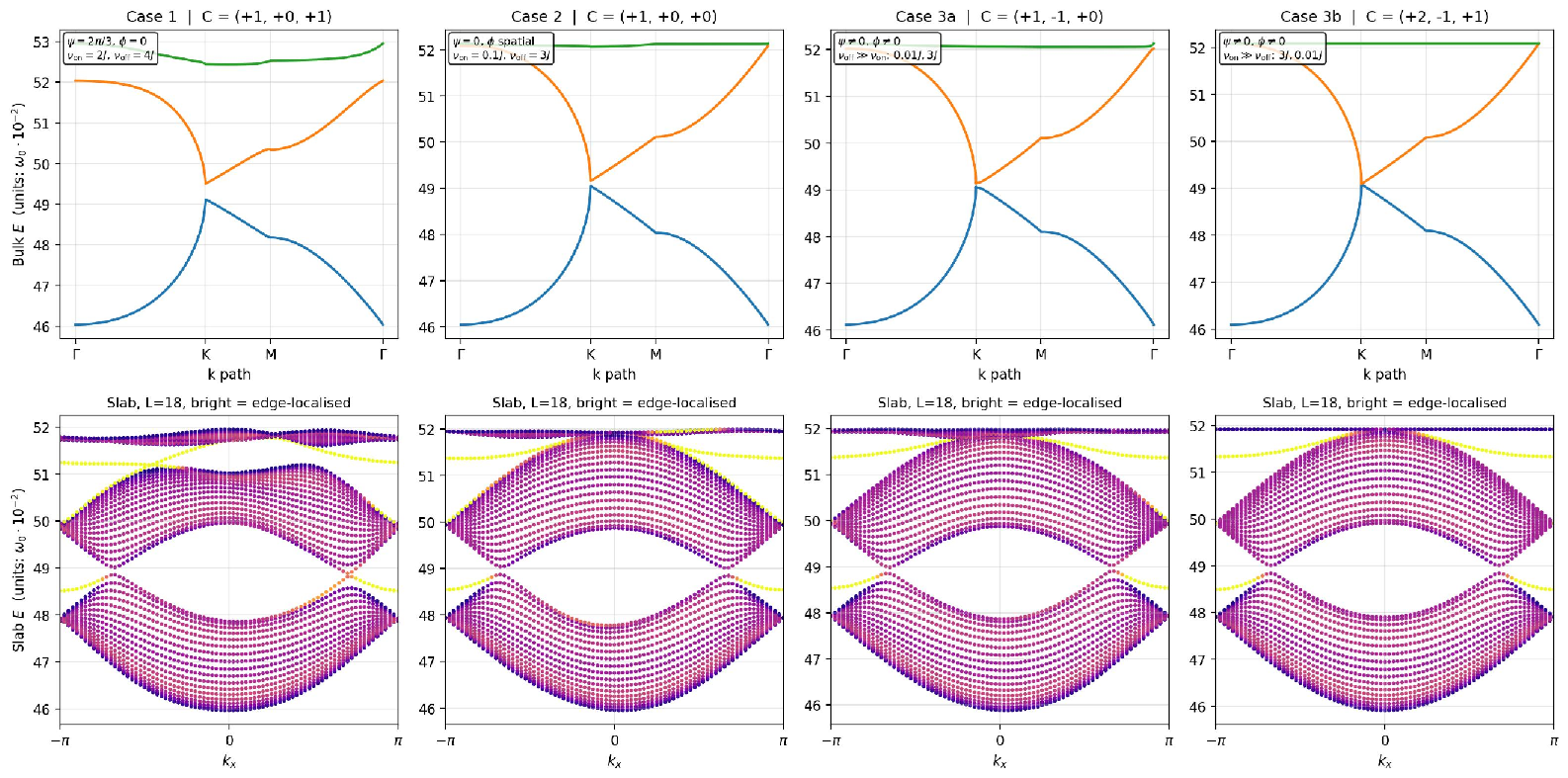}
  \caption{Bulk topological certification and zigzag-ribbon
  spectra. Top row: bulk band structure along a high-symmetry path
  in the Brillouin zone, with the para-unitary Chern number of each
  band labelled; these integer Chern numbers are the primary
  topological data and are stable under modest variations of the
  parameters (Sec.~\ref{sec:SM-VI-scaling}). Bottom row: ribbon
  spectrum on a zigzag-terminated slab of $L=18$ unit cells; each
  point is coloured by the edge-localisation weight $w(n,k_{x})$ of
  Eq.~\eqref{eq:SM-VI-edgeweight} via a perceptually-uniform
  sequential color map saturated at $w=0.6$, so that
  ``bright $=$ edge-localised'' highlights modes whose density is
  concentrated in the two outermost unit cells at each boundary,
  while dark points mark bulk-extended states. Columns correspond
  to the four representative parameter points of
  Table~\ref{tab:chern-table}: Case~1, Case~2, Case~3a and
  Case~3b. Bright (yellow) colouring indicates high $w$ and comes
  in several physically distinct varieties (band-edge folds whose
  transverse index forces support near the boundary;
  non-topological kagome boundary states; and, in bulk gaps, modes
  that carry the topological information), so high $w$ alone is
  not a topological marker.
  Across the four panels of $G_{1|2}$ this bright weight remains
  attached to the band-1 and band-2 envelopes near the projected
  Brillouin-zone edges and does not form an isolated branch that
  runs monotonically across the gap. This holds uniformly for all
  four cases, including Cases~1 and 2 with $|C_{1}|=1$: no panel
  exhibits the idealised fermionic spectral-flow signature.
  The topological certification of the four cases is therefore not
  extracted from the ribbon directly: it rests on the integer bulk
  Chern numbers labelled in the top row, which are well-defined
  para-unitary invariants and remain stable across the quadrant
  $(\nu_{\mathrm{on}},\nu_{\mathrm{off}})\in[0.005J,4J]^{2}$
  (Sec.~\ref{sec:SM-VI-scaling}), and on the chiral edge
  circulation observed in the real-space current maps of
  Figs.~1--3 of the main text and quantified by the ratio
  $\Lambda_{I}$ of Eq.~\eqref{eq:SM-IV-lambda}.
  The ribbon spectra of the bottom row provide qualitative,
  finite-$L$, momentum-space context for that certification rather
  than an independent diagnostic: in all four cases the bright
  weight that accompanies a non-zero $C_{1}$ appears in $G_{1|2}$
  as edge-localised loci attached to the band envelopes, and the
  four panels are qualitatively similar at this resolution
  irrespective of whether $|C_{1}|=1$ or $|C_{1}|=2$.
  We emphasise this point: in para-Hermitian BdG the ribbon
  morphology is not expected to encode the bulk Chern number, and
  the fact that the four panels of $G_{1|2}$ look qualitatively
  alike is consistent with this expectation rather than with a
  deficiency of the diagnostic. The bright loci are not the
  idealised chiral spectral-flow branches running monotonically
  from one envelope to the other across the gap; as quantified by
  the finite-size scaling of Fig.~\ref{fig:scaling-L}, the literal
  fermionic bulk--boundary correspondence does not transfer to
  bosonic BdG systems, and the absence of a single gap-traversing
  branch is a generic feature of the para-Hermitian
  setting~\cite{SM:Peano2016,SM:Shindou2013,SM:Chaudhary2021} rather than a
  finite-resolution artefact. The deeper, dispersion-independent
  imprint of the bulk invariants is therefore the real-space
  chiral circulation of Figs.~1--3 of the main text, which the
  bosonic system does exhibit unambiguously.}
  \label{fig:chern-ribbon}
\end{figure}

\paragraph*{Bulk--boundary correspondence.}
On a zigzag-terminated slab of $L=18$ unit cells we count the number $N_{\chiralcur}(G_{n|n+1})$  of chiral branches per edge that cross the gap $G_{n|n+1}$ between bands $n$ and $n+1$,. To distinguish edge-localised modes from bulk states in the ribbon spectrum, each BdG eigenstate $|\psi_{n,k_{x}}\rangle$ (a $6L$-component vector indexed by
$y\in\{1,\dots,L\}$, sublattice $s\in\{A,B,C\}$ and particle/hole
sector $\tau\in\{+,-\}$) is assigned a scalar
\emph{edge-localisation weight}
\begin{equation}
w(n,k_{x})
=\frac{\sum_{\alpha\in\mathcal B}\,|\psi_{n,k_{x}}(\alpha)|^{2}}
{\sum_{\alpha=1}^{6L}\,|\psi_{n,k_{x}}(\alpha)|^{2}}\in[0,1],
\label{eq:SM-VI-edgeweight}
\end{equation}
where $\mathcal B$ denotes the set of components supported on the
two outermost unit cells at each boundary, i.e.\ the four values
$y\in\{1,2\}\cup\{L-1,L\}$, summed over both sublattice and
particle/hole indices. A uniformly extended bulk state has
$w\simeq 4/L\approx 0.22$, whereas an ideal edge mode confined to the
boundary strip has $w\to 1$. In Fig.~\ref{fig:chern-ribbon} this
weight controls the colour of each ribbon eigenstate via a
perceptually-uniform sequential color map saturated at $w=0.6$, so
that dark points mark bulk-extended states and bright points mark
edge-localised modes; the caption label ``bright $=$ edge-localised''
refers to this convention.

It is important to emphasise that the weight $w(n,k_{x})$ is a purely
geometric diagnostic of boundary localisation, not a topological
marker. A ribbon eigenstate can acquire $w\gg 4/L$ for several
unrelated reasons: (i) it is a genuine chiral gap-crossing mode
implied by the bulk--boundary correspondence; (ii) it is a
non-topological boundary state of the kagome \emph{single-particle}
spectrum, which in the pairing-free tight-binding limit is identical
for fermionic and bosonic excitations and is well known to host
termination-dependent edge modes
\cite{SM:NakadaFujita1996,SM:Sekh2026,SM:Xia2020,SM:Gulevich2017}; or (iii) it is a bulk-band edge branch whose transverse index
forces its support to lie near the boundary rows. All three
contributions colour the ribbon plots bright, so high $w$ alone is
not a topological diagnostic; the relevant question is which bright
features, if any, sit inside a bulk gap and carry information about
the band Chern numbers. In a fermionic two-dimensional Chern
insulator the diagnostic would be the appearance, inside each bulk
gap $G_{n|n+1}$, of $|\sum_{m\le n}C_{m}|$ isolated chiral branches
running monotonically from one envelope to the other with a definite
sign of $\partial E/\partial k_{x}$, which would correspond to the
heuristic relation
\begin{equation}
N_{\chiralcur}(G_{n|n+1})
=\Biggl|\sum_{m\le n}C_{m}\Biggr|.
\label{eq:SM-VI-BBC}
\end{equation}
We stress at the outset that, in the para-Hermitian BdG setting
relevant here, this spectral-flow signature is \emph{not} a theorem
and is not, in fact, what one observes in any of the four cases of
Fig.~\ref{fig:chern-ribbon}, including those with $|C_{1}|=1$: the
high-$w$ loci that accompany a non-zero $C_{1}$ remain attached to
the band-1 and band-2 envelopes near the projected Brillouin-zone
edges and do not detach into a single gap-traversing branch.
Reading Fig.~\ref{fig:chern-ribbon} with this in mind, all four
panels of $G_{1|2}$ share the same qualitative morphology: bright
weight pinned to the envelopes, no isolated chiral branch in the
interior of the gap. Cases~3a ($|C_{1}|=1$) and 3b ($|C_{1}|=2$),
in particular, produce ribbon spectra that are visually
indistinguishable at this resolution, and so do Cases~1 and~2
($|C_{1}|=1$) at the same level of detail: the change of $|C_{1}|$
across the four cases is not mirrored by a more concentrated or
more intense boundary-weight pattern on the slab. We regard this as
a feature of the framework rather than a shortcoming: in
para-Hermitian BdG the ribbon morphology is not expected to encode
the bulk Chern number~\cite{SM:Peano2016,SM:Shindou2013,SM:Chaudhary2021},
and the present finding makes that prediction explicit. The chiral
information predicted by a non-zero $C_{1}$ is therefore not
recovered from the ribbon dispersion, but from two complementary
observables that the bosonic system does exhibit unambiguously:
(i)~the integer bulk Chern numbers computed from the para-unitary
Berry curvature, which are well-defined topological invariants and
remain stable across the entire quadrant
$(\nu_{\non},\nu_{\noff})\in[0.005J,4J]^{2}$
(Sec.~\ref{sec:SM-VI-scaling}); and (ii)~the chiral edge circulation
observed in the real-space current maps of Figs.~1--3 of the main
text and quantified by the phase-sensitive ratio $\Lambda_{I}$ of
Eq.~\eqref{eq:SM-IV-lambda}. The ribbon spectrum of
Fig.~\ref{fig:chern-ribbon} is therefore best read as qualitative
momentum-space support for the bulk-Chern certification, not as a
primary topological diagnostic; the deviation from the idealised
spectral-flow picture on $G_{1|2}$ is examined in detail, and at
much higher resolution, in Sec.~\ref{sec:SM-VI-scaling}. Two
physical facts stand out:

\begin{itemize}[leftmargin=*,itemsep=3pt]
\item \emph{Pairing phases alone generate topology} (Case~2).
      With $\psi=0$ the hopping sector is time-reversal symmetric,
      yet $C_{1}=+1$ and a single chiral edge mode crosses the lower
      gap. The only source of symmetry-breaking is the
      sublattice-resolved pairing phase pattern.
\item \emph{Confined regime carries $|C_{1}|=2$ in the bulk}
      (Case~3b). When the on-site pairing dominates, the lowest band
      carries Chern number $+2$, the largest first-Chern in our
      parameter set; the $L=18$ ribbon shows enhanced edge-weight
      features in the lower gap whose interpretation is refined in
      Sec.~\ref{sec:SM-VI-scaling}, where finite-size scaling
      indicates that they are pinned to the bulk envelopes rather
      than traversing $G_{1|2}$ as chiral branches at the
      resolutions accessible to us. The non-perturbative imprint of
      $|C_{1}|=2$ is instead the confined chiral transport observed
      in the real-space current maps (Fig.~3 of the main text).
\end{itemize}

In the pairing-off limit ($\nu_{\non},\nu_{\noff}\to 0^{+}$) the
algorithm recovers the standard $\pi/2$-flux kagome tight-binding
Chern sequence $(0,-1,+1)$~\cite{SM:Peano2016,SM:Chaudhary2021}, which we
have used as a quantitative benchmark of the implementation.

\subsection{Finite-size scaling in the ribbon and subtleties of
bosonic bulk--boundary correspondence}\label{sec:SM-VI-scaling}

\begin{figure}[t!]
  \centering
  \includegraphics[width=\columnwidth]{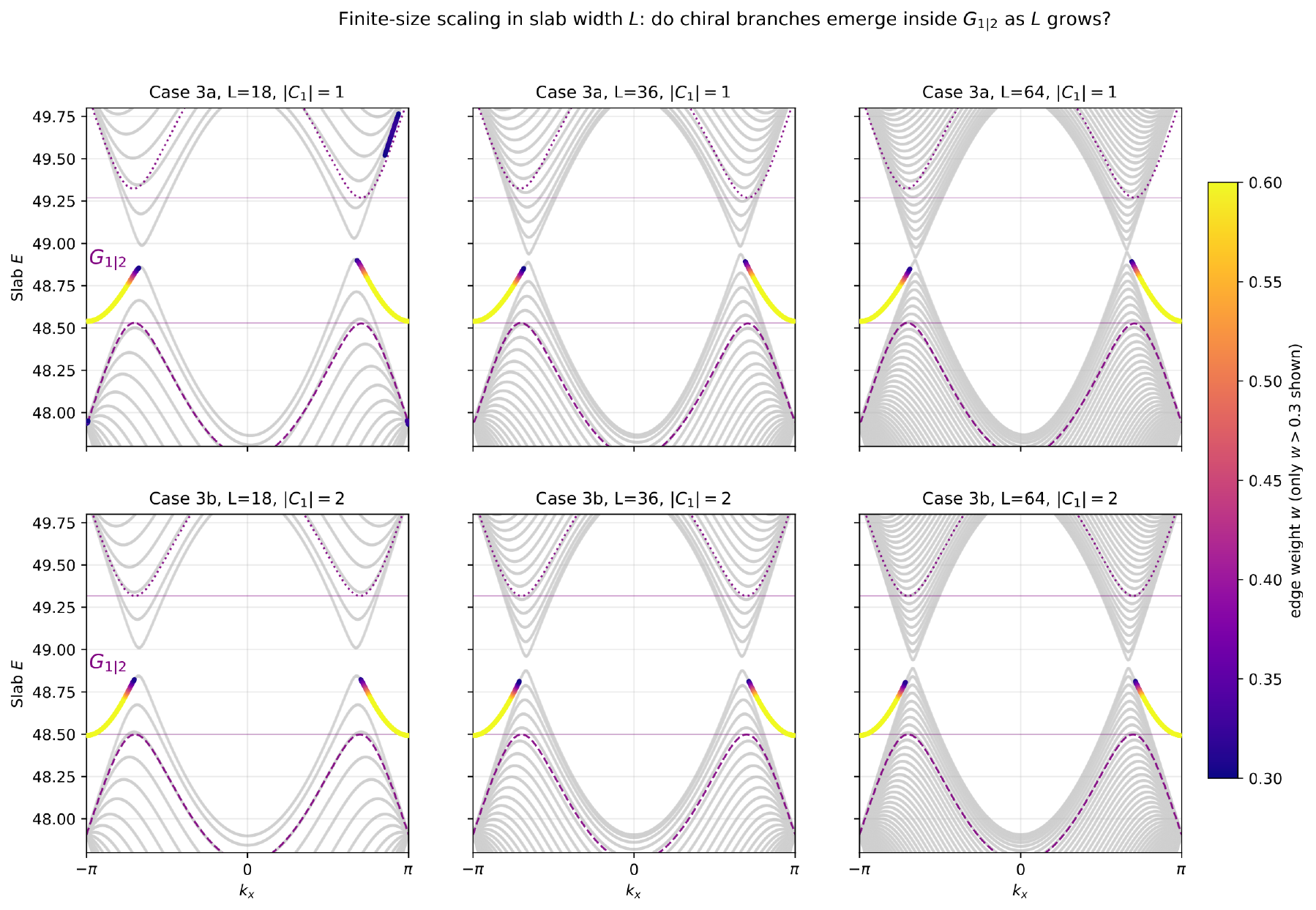}
  \caption{Finite-size scaling of the slab spectrum on the gap
  $G_{1|2}$ for Case~3a (top) and Case~3b (bottom), at
  $L\in\{18,36,64\}$ (columns) and $n_{k_{x}}=801$.
  Grey points: all para-unitary eigenvalues in the window.
  Coloured points: only modes with edge-localisation weight $w>0.30$
  (color scale: $w\in[0.30,0.60]$).
  Dashed (dotted) purple curves mark the top of band~1 (bottom of
  band~2) from a $201\times 201$ scan of the Brillouin zone; the
  interior of $G_{1|2}$ lies between them.
  For both cases, increasing $L$ from $18$ to $64$ does not populate
  the interior of $G_{1|2}$ with a chiral branch: the edge-localised
  weight remains attached to the bulk envelopes near
  $k_{x}\sim\pm 2\pi/3$ rather than detaching into the interior of
  the gap.
  The number of edge-localised modes inside the gap window of each
  panel does not grow with $L$: an independent count of points with
  $w>0.3$ that lie strictly between the band-1 and band-2 envelopes
  saturates at the same value for $L=36$ and $L=64$ in both Case~3a
  and Case~3b, indicating that the residual in-gap weight comes
  from numerical proximity to the envelope rather than from an
  emerging chiral branch.
  This is the expected behaviour in para-Hermitian BdG settings,
  where well-defined integer band Chern numbers (top row of
  Fig.~\ref{fig:chern-ribbon}) need not manifest as a single
  chiral spectral-flow branch on a slab; the present scan up to
  $L=64$, which spans nearly four times the slab width used in
  Fig.~\ref{fig:chern-ribbon} and reaches a regime in which the
  most edge-localised modes have visibly settled into a stable
  transverse profile, shows no tendency for the bright weight to
  detach from the envelopes as $L$ grows, and thereby justifies
  reading the bulk Chern numbers --- together with the chiral edge
  circulation of Figs.~1--3 of the main text and the ratio
  $\Lambda_{I}$ of Eq.~\eqref{eq:SM-IV-lambda} --- as the actual
  physical signature of topology in this bosonic system.
  We note that at
  $L=18$, top of the Case~3a panel, a faint dark-violet feature
  appears near $k_{x}\sim+\pi$ just above the dotted envelope
  ($E\!\sim\!49.5\text{--}49.8$): these are bulk band-2 states with
  finite-size edge leakage $w\!\approx\!0.30\text{--}0.32$,
  marginally above the $w>0.30$ cut. They lie \emph{outside}
  $G_{1|2}$, lie within band~2 itself, and disappear entirely already
  at $L=36$, illustrating the rapid suppression of bulk leakage as
  $\xi/L\to 0$. The same finding
  holds on $G_{2|3}$, which is additionally too narrow in Case~3b
  ($\min\Delta\sim3\cdot10^{-4}$) to carry any resolvable flow.}
  \label{fig:scaling-L}
\end{figure}

The ribbon spectra of Fig.~\ref{fig:chern-ribbon} used $L=18$ unit
cells across the slab and $n_{k_{x}}=81$ samples in $k_{x}$; this
resolution is sufficient to resolve the bulk envelopes and the
edge-localised features we highlighted, but it is coarse compared with
the scales that would be required to see a conventional chiral
spectral flow through the gap $G_{1|2}$ of Case~3. As a direct test of
whether the standard fermionic bulk--boundary correspondence picture
(an edge branch whose graph crosses $G_{1|2}$ from the top of band~1
to the bottom of band~2 as $k_{x}$ runs over $[-\pi,\pi]$) is a
quantitatively accurate description of our bosonic BdG spectra, we
repeated the slab calculation for Case~3 in two independent ways:

\begin{itemize}
\item[(i)] High angular resolution, $n_{k_{x}}=801$, with $L=24$, finding that
the features previously interpreted as ``chiral branches inside the
gap'' merge smoothly with the upper envelope of band~1, rather than
peeling off to traverse the gap.
\item[(ii)] A systematic scaling in slab
width, $L\in\{18,36,64\}$ at $n_{k_{x}}=801$, keeping only modes with
edge-localisation weight $w>0.30$.
\end{itemize}
Figure~\ref{fig:scaling-L} shows the outcome: the bright points remain attached to the projected-bulk envelopes of bands~1 and~2 and do \emph{not} detach into the interior of $G_{1|2}$ as $L$ grows; at $L=64$, for which the bulk correlation length $\xi$ is well below the slab width, a genuine chiral edge branch would have had to emerge, yet none does.

\paragraph*{Why the integer Chern indices do not enforce a rigid bulk--boundary correspondence.}
The absence of monotonic growth of the edge-localised count with $L$ in Fig.~\ref{fig:scaling-L}, together with the
integer-valued Chern numbers computed in all four cases, raises a structural question that deserves an
explicit answer. One might be tempted to attribute the breakdown of the standard bulk--boundary
correspondence (BBC) to the non-conservation of the particle number, $[H_{B},\hat N_{B}]\neq 0$, induced by
the pairing terms. This is not the correct explanation: fermionic BdG Hamiltonians share that feature
and nevertheless exhibit a rigid BBC, as established for $p$-wave superconductors and Kitaev-class
topological superfluids~\cite{SM:Kitaev2001,SM:ReadGreen2000,SM:AltlandZirnbauer1997}. The asymmetry between
the two cases is structural rather than symmetry-theoretic, and resides in the geometry of the
Bogoliubov bundle itself.

In a Nambu basis $\Psi_{B}=(a_{\mathbf{k}},a^{\dagger}_{-\mathbf{k}})^{T}$, the dynamical generator of
bosonic BdG dynamics takes the form $\mathcal{H}_{B}(\mathbf{k})=\sigma_{z}H_{B}(\mathbf{k})$, with $H_{B}$ Hermitian
and $\sigma_{z}=\mathrm{diag}(\mathbf{1}_{m},-\mathbf{1}_{m})$ the indefinite metric of the particle--hole sectors,
where $m$ denotes the number of physical bands per unit cell of the lattice (so that the BdG matrix
is of size $2m\times 2m$). For the kagome lattice considered in this work, $m=3$.
Within the regime considered here, in which $H_{B}(\mathbf{k})$ is positive-definite on the
entire Brillouin zone, $\mathcal{H}_{B}$ is brought to diagonal form by a \emph{para-unitary}
transformation $T$ satisfying $T^{\dagger}\sigma_{z}T=\sigma_{z}$, and its spectrum is real and arranged in
particle--hole conjugate pairs $\pm\omega_{n}(\mathbf{k})$. The set of admissible $T$'s does not form the
compact unitary group $U(2m)$ that controls the fermionic BdG case, but rather the \emph{non-compact
para-unitary group} $U(m,m)$ of linear maps preserving the indefinite metric
$\sigma_{z}$~\cite{SM:Mostafazadeh2002,SM:Shindou2013,SM:KondoAkagiKatsura2020}.
This change of structure group has direct topological consequences.

In the fermionic BdG case, where the structure group is the compact $U(2m)$, the band Chern integer
\begin{equation}
C_{n}=\frac{1}{2\pi i}\int_{\mathrm{BZ}}\mathrm{Tr}\!\left[\,P_{n}\,dP_{n}\wedge dP_{n}\,\right]
\end{equation}
counts, by an Atiyah--Singer-type index theorem applied to the slab boundary, the number of chiral
edge channels that traverse the gap with a definite sign of $\partial E/\partial k_{x}$. Two ingredients
of that theorem are essential: (a) the spectral projector $P_{n}$ is orthogonal with respect to a
positive-definite inner product, so that the bulk--boundary index is a difference of \emph{positive}
quantities; and (b) the structure group $U(2m)$ is compact, ensuring that the boundary index is a
homotopy invariant of the bulk Hamiltonian. In the bosonic case the natural projector
$P_{n}=T_{n}\sigma_{z}T_{n}^{\dagger}\sigma_{z}$ (obtained from the para-unitary inverse
$T^{-1}=\sigma_{z}T^{\dagger}\sigma_{z}$ implied by the defining condition $T^{\dagger}\sigma_{z}T=\sigma_{z}$,
and selecting the columns of $T$ associated with band $n$; see~\cite{SM:KondoAkagiKatsura2020} for the
full derivation) is orthogonal only with respect to the indefinite metric $\sigma_{z}$, in the
sense that $\sigma_{z}P_{n}^{\dagger}\sigma_{z}=P_{n}$ rather than $P_{n}^{\dagger}=P_{n}$; and the
para-unitary group $U(m,m)$ is non-compact. The bulk Chern integer remains
well-defined and gauge-invariant under $U(m,m)$ transformations, and we have verified numerically that
it takes integer values across the four cases. However, the boundary index obtained by repeating the
fermionic construction is no longer a difference of positive quantities but a \emph{signed} sum over
boundary modes weighted by their $\sigma_{z}$-norm sign, and the non-compactness of $U(m,m)$ removes the
homotopy invariance that would have anchored that signed sum to $C_{n}$. The Chern integer therefore
ceases to count protected chiral edge channels even when the bulk gap remains open and the bulk
invariant is unambiguously non-zero~\cite{SM:Mostafazadeh2002,SM:Shindou2013,SM:KondoAkagiKatsura2020}.

The bosonic Chern index is accordingly best read as a \emph{phase-sensitive observable} of the bulk
Bogoliubov bundle, certified by the integer values reported in the top row of Fig.~\ref{fig:chern-ribbon} and stable
across the quadrant $(\nu_{\non},\nu_{\noff})\in[0.005J,4J]^{2}$ (Sec.~\ref{sec:SM-VI-scaling}), rather than as a strict
counter of protected edge modes. This is exactly the picture extracted from Fig.~\ref{fig:scaling-L}: the
edge-localised count saturates at a finite, $L$-independent value because the modes responsible for it
are not anchored by a fermionic index theorem, and the dispersion-independent topological signature in
this bosonic system is the chiral edge circulation of Figs.~1--3 of the main text, quantified by the
ratio $\Lambda_{\cal I}$ of Eq.~\eqref{eq:SM-IV-lambda}.

We interpret this observation as a manifestation of a well-known
subtlety: the fermionic bulk--boundary correspondence does not
transfer literally to bosonic Bogoliubov--de Gennes systems. The
para-unitary Hamiltonian $\Sigma_{z}H_{k}$ is not
Hermitian but para-Hermitian; its band-resolved Chern numbers are
well-defined integers, and they remain stable under smooth parameter
variations (we have verified this numerically across the entire
quadrant $(\nu_{\non},\nu_{\noff})\in[0.005J,4J]^{2}$), but the
corresponding edge states of a half-infinite ribbon need not appear
as a single chiral spectral-flow branch in the band gap. Physically,
the BdG bosonic quasiparticles mix particle and hole sectors with
opposite paravector metric, and the ``edge mode'' demanded by a
non-zero $C_{1}$ can be distributed among several quasiparticle
branches, can be admixed with the bulk continuum near the band edges,
or can be pinned to the boundary at energies outside the gap~---~
behaviours all consistent with the spectra above and with the general
analyses of Peano \emph{et al.}~\cite{SM:Peano2016},
Shindou~\cite{SM:Shindou2013}, and Chaudhary \emph{et
al.}~\cite{SM:Chaudhary2021}.

A second mechanism, complementary to the $\xi/L$ argument above,
further hampers the slab analysis specifically in Case~3b: the
uppermost BdG band is significantly compressed (bandwidth
$W_{3}/W_{1}\approx 0.20$ at $L=64$), reminiscent of the well-known
flat band of the pristine kagome single-particle spectrum that is
only weakly dispersed by the pairing and Zeeman terms. As a
consequence, $G_{2|3}$ in Case~3b is anomalously narrow
($\min\Delta\sim 3\cdot 10^{-4}$) and the $L$-fold quasi-degenerate
set of states confined within the narrow bandwidth $W_{3}$ acts as
a near-degenerate manifold whose individual eigenvectors are
sensitive to the conditioning of the para-unitary problem. This
compounds the finite-size limitation: even at $L=64$, $G_{2|3}$ is
too narrow and the band-3 manifold too compressed to host a
resolvable chiral branch, regardless of the underlying topology.
The bulk Chern indicator, the flake-circulation diagnostic, and the
leakage ratio $\Lambda_{\cal I}$ are unaffected by this compression and
therefore remain the appropriate certificates.

Far from being a limitation specific to our analysis, this is a
generic feature of bosonic pairing topology that the present work
turns into an asset: the real-space currents, computed
non-perturbatively on finite flakes, are more informative and more
reliable diagnostics of the underlying topology than the ribbon
spectral-flow argument, which has no literal counterpart in the
bosonic BdG setting. Accordingly, our evidence for non-trivial
chiral transport \emph{does not rely on a ribbon-BBC spectral-flow
argument}. It rests on three independent and mutually reinforcing
pillars:

\begin{itemize}
\item[(a)] fully non-perturbative simulations on finite
zigzag-terminated flakes (main text, Figs.~1--3), which exhibit
chiral circulation of both $\langle \cal J \rangle$ and
$\langle \cal I \rangle$, including the case $\psi=0$ with only
pairing phase present;
\item[(b)]integer, gauge-invariant, stable para-unitary Chern numbers of the
bulk bands, computed with the Fukui--Hatsugai--Suzuki prescription (Tab.~\ref{tab:chern-table}), which undergo a discrete reorganisation
$(+1,-1,0)\to(+2,-1,+1)$ between the two regimes of Case~3 while the
bulk gap stays open along a smooth path in parameter space;
\item[(c)] the phase-sensitive leakage ratio $\Lambda_{\cal I}$ of
Eq.~\eqref{eq:SM-IV-lambda} (with its pointwise counterpart
$\mathcal{R}=|\langle \cal I\rangle/\langle \cal J \rangle|$ at
a defect site), which varies continuously with $(\psi,\varphi)$
and with $\nu_{\non}/\nu_{\noff}$ across the entire parameter
plane and whose anomalous enhancement is inaccessible to any
particle-conserving model.
\end{itemize}
We stress that, unlike the ribbon-BBC spectral-flow argument, each of (a)--(c) survives the subtleties of bosonic BdG described above.

\section{Leakage ratios: systematic scan}\label{sec:SM-VII}

To provide a complete view of the anomalous-scattering signature, we
sweep the two BP couplings $(\nu_{\non},\nu_{\noff})$ on a uniform
(linear) grid covering the physical range
$\nu_{\non}\in[0,10J]$, $\nu_{\noff}\in[3J,5.5J]$ for Case~1 and
$\nu_{\non},\nu_{\noff}\in[0,3J]$ for Case~3, and compute the
leakage ratios $\Lambda_{\cal J}$ and
$\Lambda_{\cal I}$ [Eq.~\eqref{eq:SM-IV-lambda}] on the same 14-site loop $\mathcal L$ surrounding a connected edge-defect cluster
$\{\mathcal S\}$, for Case~1 (uniform hopping phase) and Case~3 (both
sectors dressed). For each grid point we evaluate the steady-state
currents from the para-unitary eigenvector of the BdG matrix closest
to the operating frequency $\omega_{0}$. We observe the following qualitative features:
\begin{itemize}[leftmargin=*,itemsep=3pt]
\item In Case~1, $\Lambda_{\cal J}$ remains small over most of the plane
      and is enhanced only in a narrow diagonal ridge corresponding
      to the inter-loop eigenmode discussed in
      Sec.~\ref{sec:SM-IVB}. $\Lambda_{\cal I}$ is likewise modest.
\item In Case~3 both ratios exhibit a sharp crossover across the
      diagonal $\nu_{\non}=\nu_{\noff}$: $\Lambda_{\cal I}$ is large in
      the anomalous regime $\nu_{\noff}\gg\nu_{\non}$ (upper-left
      quadrant), where the PC channel leaks into the bulk around the
      defect; $\Lambda_{\cal I}$ collapses in the confined regime
      $\nu_{\non}\gg\nu_{\noff}$ (lower-right quadrant), which is
      topologically protected by $|C_{1}|=2$ (Sec.~\ref{sec:SM-VI}).
      The map of $\Lambda_{\cal J}$ is complementary but weaker.
\end{itemize}
These two trends reproduce the central claim of the main text: the
anomalous leakage carried by the pairing current is a genuinely new
observable, absent in any particle-conserving theory, and is tunable
through the BP phase structure and the $\nu_{\non}/\nu_{\noff}$
ratio. The corresponding maps of $\Lambda_{\cal J}$ and
$\Lambda_{\cal I}$ on the $(\nu_{\non},\nu_{\noff})$ plane are
reproduced in Fig.~4 of the main text.

\section{Experimental outlook: extended analysis}\label{sec:SM-VIII}

The two predictions that most directly test our central claim,
reformulated in light of the finite-size scaling analysis of
Sec.~\ref{sec:SM-VI-scaling}, are:
\begin{enumerate}
\item[(a)] Chiral circulation of the pairing current
      $\langle \mathcal{I}\rangle$ along the edge of a finite
      zigzag-terminated sample at vanishing hopping phase ($\psi=0$)
      but non-trivial pairing phase pattern. This is a direct
      consequence of the open-boundary Hamiltonian and does not
      require a ribbon spectral-flow argument.
\item[(b)] A qualitative reorganisation of the chiral current pattern
      between the anomalous ($\nu_{\noff}\gg\nu_{\non}$) and confined
      ($\nu_{\non}\gg\nu_{\noff}$) regimes of Case~3, reflected in
      both the spatial profile of $\langle \mathcal{I}\rangle$ on a
      flake and in the anomalous enhancement of the phase-sensitive
      ratio $\mathcal{R}=|\langle \mathcal{I}\rangle/\langle
      \mathcal{J}\rangle|$ at a localised defect; this transition is
      consistent with the bulk Chern reorganisation
      $(+1,-1,0)\to(+2,-1,+1)$.
\end{enumerate}
We emphasise two points of scientific transparency.
First, we do \emph{not} predict an integer jump
``one chiral edge branch $\to$ two chiral edge branches'' in a
half-infinite-ribbon transport experiment: as shown in
Fig.~\ref{fig:scaling-L}, such a sharp spectral-flow signature is
not supported by our slab calculations at $L$ up to $64$, a
consequence of the subtleties of bulk--boundary correspondence in
bosonic BdG systems (Sec.~\ref{sec:SM-VI-scaling}).
Second, the observable $\mathcal{R}$, being a dimensionless local
ratio at a defect, is calibration-free, survives the bulk-boundaru correspondence subtleties, and constitutes the primary quantitative signature of the pairing-enabled chiral transport we report.

\subsection{Bosonic nature of the platform}\label{sec:SM-VIII-bosonic}

Before describing the implementation it is worth emphasising that
the Hamiltonian of Eq.~(1), although conventionally assigned to
``circuit-QED'', lives in the \emph{resonator}, not the qubit, branch
of that platform. Superconducting circuit architectures host two
complementary families of elements: (i) nonlinear two-level systems
(transmons, fluxoniums, flux qubits) whose excitations are hard-core
and therefore effectively fermionic, and (ii) linear microwave
resonators (LC circuits, coplanar-waveguide cavities, 3D cavities),
whose single-mode excitations are indistinguishable bosons obeying
$[a,a^{\dagger}]=1$ with unrestricted occupation. Our lattice sites
$(j,s)$ are modes of the second type: each is a harmonic LC resonator
of frequency $\omega_{0}=1/\sqrt{LC}$, and all creation and
annihilation operators in Eq.~(1) are standard bosonic
$a_{j,s}, a_{j,s}^{\dagger}$. In this sense the proposal is closer
in spirit to the photonic and polaritonic
implementations~\cite{SM:Hafezi2013,SM:Hafezi2011,SM:Gulevich2017} than to
qubit-based superconducting-circuit quantum simulators; microwaves
simply replace optical photons as the carrier. Consequently, the
BdG machinery of Sec.~\ref{sec:SM-I}---para-unitary diagonalisation, commutation with $\Sigma_{z}$, non-conservation of the total photon number---is not an artefact of the lattice model but the physically correct
framework for a linear array of parametrically driven bosonic
resonators.

The three terms of Eq.~(1) are then engineered as follows.
The on-site term $\omega_{0}a_{j,s}^{\dagger}a_{j,s}$ is the bare
LC energy. The hopping $J\,e^{i\psi_{s,s'}}a_{j,s}^{\dagger}a_{j',s'}$
is a particle-conserving beam-splitter interaction, induced either
by direct capacitive/inductive coupling together with a modulated
drive or, more flexibly, by a flux-pumped SQUID inserted between
neighbouring resonators~\cite{SM:Pocklington2023}: the
time-dependent Josephson inductance produced by a SQUID flux
$\Phi_{\mathrm{ext}}(t)=\Phi_{\mathrm{dc}}+\Phi_{\mathrm{ac}}
\cos(\omega_{m}t+\psi_{s,s'})$ generates, within RWA, a Peierls
phase on the corresponding lattice bond with $\psi_{s,s'}$
inherited directly from the drive phase.
The bosonic-pairing terms
$\nu_{\non}e^{i\phi_{s}}(a_{j,s}^{\dagger})^{2}$ and
$\nu_{\noff}e^{i\phi_{s,s'}}a_{j,s}^{\dagger}a_{j',s'}^{\dagger}$
are, respectively, one- and two-mode microwave squeezing
interactions, the canonical output of degenerate and non-degenerate
parametric amplifiers based on Josephson junctions or SQUIDs driven
at $2\omega_{0}$~\cite{SM:Pocklington2023,SM:Yurke1989,SM:Flurin2012}.
Amplitudes $\nu_{\non,\noff}$ and phases $\phi_{s},\phi_{s,s'}$
track the amplitudes and phases of the classical $2\omega_{0}$ pumps
acting on each resonator and on each bond, respectively.
Particle-number non-conservation, intrinsic to our model, is
physically supplied by these classical pumps, which act as
reservoirs of photon pairs; they do not introduce fermionic
statistics at any step.

\subsection{Primary platform: circuit-QED parametric
arrays}\label{sec:SM-VIII-cQED}

The most direct realisation is a circuit-QED array of parametrically
driven superconducting resonators in the kagome geometry, along the
lines of Pocklington, Wang, and Clerk~\cite{SM:Pocklington2023}:
nearest-neighbour hopping $J$ is set by capacitive or inductive
coupling with a flux-tunable phase $\psi$ threaded through SQUID
loops, while the on-site and off-diagonal pairing amplitudes
$\nu_{\non}$ and $\nu_{\noff}$---together with the pairing phase
$\varphi$---are generated by two independent parametric pumps at
frequency $2\omega_{0}$: one modulating each resonator and one
modulating the coupling elements.

With currently reported parameters ($J/2\pi\sim 10$--$50$~MHz,
parametric couplings up to $\sim 0.3\,J$, internal losses
$\kappa/J\lesssim 10^{-2}$), the four regimes of
Table~\ref{tab:chern-table} map to experimentally distinct operating
points. Case~2 is reached by turning off the SQUID flux and keeping
only the parametric-pump phase pattern; Case~3b is reached by
increasing the local parametric drive until
$\nu_{\non}/\nu_{\noff}\gtrsim 10$.

\subsection{Discriminating observables}\label{sec:SM-VIII-obs}

The observables that cleanly discriminate the pairing-induced channel
from any particle-conserving one are, in descending order of
robustness:
\begin{enumerate}[leftmargin=*,itemsep=3pt]
\item \textbf{Primary observable.} The ratio of pairing to hopping
      current at a single defect site,
      $\mathcal{R}=|\langle \mathcal{I}\rangle/\langle \mathcal{J}\rangle|$,
      whose anomalous enhancement is the defect-scattering signature
      predicted here. Being a dimensionless local ratio, $\mathcal{R}$
      is calibration-free and extracts the pairing-induced signal
      without requiring absolute photon-number normalisation.
      $\mathcal{R}$ is well-defined for any Hamiltonian phase choice,
      is bounded in each bulk-gapped phase, and attains a
      sign-definite anomalous enhancement in the
      $\nu_{\noff}\gg\nu_{\non}$ regime.
      This is the killer observable of the proposal.
\item \textbf{Secondary observable.} The site-resolved steady-state
      photon current $\langle \mathcal{I}\rangle(\mathbf{r})$ under a
      weak coherent probe at one edge of a finite zigzag-terminated
      array, mapped in real space. This observable registers the
      chiral circulation of the pairing current along the finite
      edge and distinguishes the anomalous and confined regimes
      through a qualitative change of the spatial pattern (as
      illustrated in Fig.~3 of the main text). It is not an integer
      mode count and should not be interpreted as such.
\end{enumerate}

\subsection{Complementary platforms}\label{sec:SM-VIII-alt}

Equivalent tests are available in photonic lattices with
parametric-OPO couplings along the lines of Hafezi
\emph{et al.}~\cite{SM:Hafezi2013,SM:Hafezi2011}, and in exciton-polariton
lattices with parametric scattering~\cite{SM:Gulevich2017}, where the
two currents can be read out, respectively, from
intensity-correlation measurements on the edge waveguides and from
momentum-resolved photoluminescence.